\documentclass[reprint,amsmath,amssymb,pra,superscriptaddress]{revtex4-1}
\usepackage{graphicx}
\usepackage{bm}
\usepackage{hyperref}
\hypersetup{breaklinks=true}
\usepackage[capitalise]{cleveref}

\begin{document}
\title{Generalized Theory of Optical Resonator and Waveguide\\Modes and their Linear and Kerr Nonlinear Coupling}

\author{Jonathan~M.~Silver}
\email{jonathan.silver@npl.co.uk}
\affiliation{National Physical Laboratory, Hampton Road, Teddington TW11 0LW, UK}
\author{Pascal~Del'Haye}
\email{pascal.delhaye@mpl.mpg.de}
\affiliation{Max Planck Institute for the Science of Light, Staudtstrasse 2, 91058 Erlangen, Germany}
\affiliation{Department of Physics, Friedrich--Alexander University Erlangen--Nuremberg, 91058 Erlangen, Germany}

\date{\today}

\begin{abstract}
We derive a general theory of linear coupling and Kerr nonlinear coupling between modes of dielectric optical resonators from first principles. The treatment is not specific to a particular geometry or choice of mode basis, and can therefore be used as a foundation for describing any phenomenon resulting from any combination of linear coupling, scattering and Kerr nonlinearity, such as bending and surface roughness losses, geometric backscattering, self- and cross-phase modulation, four-wave mixing, third-harmonic generation and Kerr frequency comb generation. The theory is then applied to a translationally symmetric waveguide in order to calculate the evanescent coupling strength to the modes of a microresonator placed nearby, as well as the Kerr self- and cross-phase modulation terms between the modes of the resonator. This is then used to derive a dimensionless equation describing the symmetry-breaking dynamics of two counterpropagating modes of a loop resonator and prove that cross-phase modulation is exactly twice as strong as self-phase modulation only in the case that the two counterpropagating modes are otherwise identical.
\end{abstract}

\maketitle

\section{Introduction}\label{sec:intro}

Since research into dielectric optical microcavities and microresonators began in the late 1980s~\cite{braginskQualityFacto1989,vahalaOpticalMicr2003}, we have understood them using coupled mode theory~\cite{johnsonTheoryMorph1993,marcuseCoupledMode1985,stofferCylindricalIntegrated2005}, a framework that was first established in the 1950s in the context of waveguides~\cite{schelkunConversionM1955, yarivCoupledmode1973, hausCoupledmode1987, hausCoupledmode1991}. This approach underpins our descriptions of linear coupling between resonators and other dielectric bodies such as prisms, waveguides and tapered optical fibers~\cite{littleAnalyticThe1999,spillaneIdealityFib2003} as well as optomechanical, Brillouin and Raman coupling~\cite{aspelmeyerCavityOptomechanics2014,bahlStimulatedOptomechanical2011,kippenbeUltralowthre2004} and second- and third-order (Kerr) nonlinear optical effects~\cite{ilchenkoNonlinearOp2004,linNonlinearPh2017} including frequency comb generation~\cite{chemboKerrOptical2016,delhayeOpticalFreq2007,kippenbeMicroresonat2011}. For the latter, the modal expansion approach~\cite{chemboModalExpans2010} forms the basis of a description based on the Lugiato-Lefever equation (LLE)~\cite{lugiatoSpatialDissipative1987,chemboSpatiotemporalLugiatoLefever2013,coenModelingOctavespanning2013} that has been particularly successful in modelling soliton comb generation~\cite{herrTemporalSol2013,kippenbergDissipativeKerr2018,zhangSubmilliwattlevelMicroresonator2019}.

Another interesting effect of the Kerr nonlinearity in whispering-gallery-mode (WGM), ring and other loop microresonators is symmetry breaking between counterpropagating light~\cite{caoExperimentalDemonstration2017,delbinoSymmetryBreaking2017}, obtained for example by pumping a WGM microresonator bidirectionally via a single tapered optical fiber. Universal behaviors at the critical point of this symmetry-breaking regime~\cite{woodleyUniversalSymmetrybreaking2018,silverCriticalDyn2019} similar to those found at exceptional points~\cite{chenExceptionalPoints2017,laiObservation2019} have been demonstrated in a nonlinear enhanced gyroscope~\cite{silverNonlinearEn2021a}, and could enable other enhanced sensors e.g.\ for refractive index changes~\cite{wangNonlinearMicroresonator2015}. Meanwhile, the bistable symmetry-broken regime has been used to realise optical isolators and circulators~\cite{delbinoMicroresonat2018}, memories~\cite{delbinoOpticalMemo2021} and logic gates~\cite{moroneyLogicGates2020}. 

The symmetry breaking between counterpropagating light relies upon a well-known factor of 2 between the coefficients of Kerr cross-phase modulation (XPM) and self-phase modulation (SPM)~\cite{kaplanEnhancementSagnac1981,ghalanosKerrNonlinearityInducedModeSplitting2020}, that is also instrumental in frequency comb generation and other Kerr-nonlinearity-related bistabilities, multistabilities and oscillatory and chaotic dynamical behaviors~\cite{kaplanOpticalBistability1981, kivsharSelffocusingTransverse2000, yangCounterpropagatingSolitons2017, joshiCounterrotat2018, baoOrthogonally2019, fatomePolarization2020, garbinAsymmetricBalance2020, garbinDissipativePolarization2021, hillEffectsSelf2020, woodleySelfSwitchingKerr2021, hillBreathingDy2020, xuSpontaneous2021}.

Whereas the coupled mode theory of waveguides is very well developed~\cite{hausCoupledmode1991}, this is less the case for microresonators, where much of the literature relates to specific geometries such as plane-wave cavities~\cite{marcuseCoupledMode1985}, microspheres~\cite{johnsonTheoryMorph1993,littleAnalyticThe1999}, microtoroids~\cite{chemboModalExpans2010} and ring waveguides~\cite{stofferCylindricalIntegrated2005}. Here we adopt a general approach that makes no assumptions about the geometry of the resonator, initially defining modes simply as basis states for the electromagnetic field and only subsequently stating a condition for them to be stationary or nearly stationary states. We then derive a Schr\"{o}dinger-like equation (equivalent to the single-photon Schr\"odinger equation) for the evolution of the amplitudes of a collection of modes under linear coupling. The treatment, given in \cref{sec:rmc,sec:wrc}, is self-contained and based entirely on Maxwell's equations, avoiding variational approaches and making approximations only when absolutely necessary, whereupon they are clearly stated. The relationship between standing- and travelling-wave modes is elucidated, as is the physical meaning of the complex amplitude of a mode. Our approach also explains why, even for a purely classical treatment, it makes sense to choose a normalisation in which the modulus-squared of the complex amplitude of a mode is proportional to the number of photons in it, rather than for example its energy. The same formalism is used to describe the modes of both resonators and waveguides as well as evanescent coupling between the two~\cite{yarivUniversalRe2000}. It can equally be applied to other linear coupling phenomena such as coupling between two resonators or two waveguides, Rayleigh backscattering~\cite{gorodetsRayleighSca2000} and bending~\cite{gorodetsEigenfrequen2007} and scattering losses~\cite{laceyRadiationLo1990, littleEstimatingS1996} in resonators and fibres, mode splitting in ring resonators caused by coupling to a waveguide~\cite{ctyrokyDualResonan2006}, and scattering of plane waves by dielectric bodies~\cite{barberScatteringE1975}.

In \cref{sec:knl} we introduce the Kerr nonlinearity into this framework, again defining everything from first principles and keeping the treatment initially very general. We briefly discuss the different phenomena that the Kerr effect gives rise to including third-harmonic generation and four-wave mixing, before focusing on SPM and XPM in the context of symmetry breaking between two modes of the same microresonator. We show that when these modes are counterpropagating but otherwise identical, the coefficient of XPM is exactly twice that of SPM, while if they are from the same WGM family but of different azimuthal order, or if they are of opposite circular polarizations but otherwise identical, XPM is slighly less than twice as strong as SPM. To the authors' knowledge, this is the first explicit demonstration that the ratio between XPM and SPM is exactly 2 for all pairs of counterpropagating but otherwise identical travelling-wave modes, independent of the geometry of the resonator.

Finally we put everything together to derive the dimensionless equations (\cref{dynEq}) that govern the evolution of the complex amplitudes of two counterpropagating modes in a microresonator pumped via a waveguide, in the presence of the Kerr nonlinearity. These equations form a basis for explaining the aforementioned symmetry breaking between counterpropagating light in WGM microresonators and the interesting dynamics associated with it~\cite{caoExperimentalDemonstration2017,delbinoSymmetryBreaking2017,woodleyUniversalSymmetrybreaking2018,silverCriticalDyn2019}. The generality of the framework developed means that it can also act as the foundation for explaining any phenomenon involving linear and Kerr nonlinear coupling in dielectric bodies. For example, it could be applied to a WGM family to derive the LLE~\cite{lugiatoSpatialDissipative1987,chemboSpatiotemporalLugiatoLefever2013,coenModelingOctavespanning2013}, which can be used to model Kerr frequency comb generation~\cite{herrTemporalSol2013,kippenbergDissipativeKerr2018,zhangSubmilliwattlevelMicroresonator2019}.

\section{Resonator modes and couplings}\label{sec:rmc}

A system of dielectric bodies surrounded by free space can be described by a spatially dependent permittivity $\varepsilon\!\left(\mathbf{r}\right)$, which we will treat for conciseness as though it is differentiable everywhere. Working in the Weyl gauge in which the scalar potential is set to zero, the optical electromagnetic field can be described purely by the vector potential $\mathbf{A}\!\left(\mathbf{r},t\right)$, which, in the absence of free charge and current, obeys the following form of Maxwell's equations:
\begin{equation}
\nabla\!\times\!\left(\nabla\!\times\!\mathbf{A}\right) = -\mu_0\,\varepsilon\frac{\partial^2\mathbf{A}}{\partial t^2}\label{AMaxwell}
\end{equation}
where $\mu_0$ is the permeability of free space. There is the additional 
constraint $\nabla\cdot\left(\varepsilon\mathbf{A}\right) = 0$~\cite{drummondQuantumTheo2014}, although for optical fields this is already implied by \cref{AMaxwell} due to the divergence-free nature of the form on its left-hand side. It is useful to describe the physics in terms of the time-evolution of complex amplitudes $\alpha_\sigma$ of a complete basis of spatial modes with vector potential profiles $\mathbf{a}_\sigma(\mathbf{r})$, which may be either real or complex, by expanding out $\mathbf{A}(\mathbf{r},t)$ as~\cite{drummondQuantumTheo2014}
\begin{equation}
\mathbf{A}(\mathbf{r},t) = \sum\limits_\sigma\left(\alpha_\sigma(t)\mathbf{a}_\sigma(\mathbf{r}) + \alpha_\sigma^*(t)\mathbf{a}_\sigma^*(\mathbf{r})\right).\label{Acplxexp}
\end{equation}
If the basis states are stationary states of the system, i.e.\ states where all fields oscillate at a single frequency, then a real basis ($\mathbf{a}_\sigma(\mathbf{r}) = \mathbf{a}^*_\sigma(\mathbf{r})$) would correspond to standing-wave modes in which the electric field vanishes everywhere twice during each period of oscillation, whereas a complex basis would correspond to modes in which different polarizations or spatial regions oscillate out of phase with each other.

In order to develop a unique and physically meaningful definition for the complex amplitudes $\alpha_\sigma$, we will start by working in a real basis $\{\mathbf{a}'_\rho(\mathbf{r})\}$ with real amplitudes $\{u_\rho(t)\}$:
\begin{equation}
\mathbf{A}(\mathbf{r},t) = 2\sum\limits_\rho u_\rho(t)\mathbf{a}'_\rho(\mathbf{r}).\label{Arealexpu}
\end{equation}
Substituting this into \cref{AMaxwell}, taking the dot product with $\mathbf{a}'_\rho$ and integrating over all space gives us
\begin{equation}
\sum\limits_{\rho'}N'_{\rho\rho'}\frac{\mathrm{d}^2u_{\rho'}}{\mathrm{d}t^2} = -\sum\limits_{\rho'}D'_{\rho\rho'}u_{\rho'}\label{urho2ode}
\end{equation}
where
\begin{gather}
D'_{\rho\rho'} = \frac{1}{\mu_0}\int\mathbf{a}'_\rho(\mathbf{r})\cdot\nabla\!\times\!\left(\nabla\!\times\!\mathbf{a}'_{\rho'}(\mathbf{r})\right)\mathrm{d}^3\mathbf{r}\\
\text{and}\quad N'_{\rho\rho'} = \int\varepsilon(\mathbf{r})\,\mathbf{a}'_\rho(\mathbf{r})\cdot\mathbf{a}'_{\rho'}(\mathbf{r})\,\mathrm{d}^3\mathbf{r}.
\end{gather}
Note that $D'_{\rho\rho'} = D'_{\rho'\rho}$ and $N'_{\rho\rho'} = N'_{\rho'\rho}$, the first of which is easy to verify via integration by parts given a suitable boundary condition at infinity. We now transform \cref{urho2ode} into two first-order differential equations by defining
\begin{equation}
v_\rho = \sum\limits_{\rho'}N'_{\rho\rho'}\frac{\mathrm{d}u_{\rho'}}{\mathrm{d}t} 
\end{equation}
such that
\begin{equation}
\frac{\mathrm{d}u_{\rho}}{\mathrm{d}t} = \sum\limits_{\rho'}\left(N'^{-1}\right)_{\rho\rho'}v_{\rho'}\quad\text{and}\quad\frac{\mathrm{d}v_{\rho}}{\mathrm{d}t} = -\sum\limits_{\rho'}D'_{\rho\rho'}u_{\rho'}.\label{uandvrho1ode}
\end{equation}
Defining the complex amplitudes $\{\alpha'_\rho=u_\rho+iv_\rho\}$, we obtain
\begin{equation}
\frac{\mathrm{d}\alpha'_\rho}{\mathrm{d}t} = -i\sum\limits_{\rho'}\left(S'_{\rho\rho'}\alpha'_{\rho'} + T'_{\rho\rho'}\alpha'^*_{\rho'}\right)\label{dalphaprimedt}
\end{equation}
where the matrices
\begin{equation}
S' = \frac{D'+N'^{-1}}{2}\quad\text{and}\quad T' = \frac{D'-N'^{-1}}{2}
\end{equation}
are real and symmetric.

We can now transform these results back into the complex basis $\{\mathbf{a}_\sigma(\mathbf{r})\}$ as long as the two bases are related by a unitary transformation:
\begin{equation}
\mathbf{a}_\sigma = \sum\limits_\rho U_{\sigma\rho}\,\mathbf{a}'_\rho\quad\text{where}\quad U^{-1}=U^\dagger.
\end{equation}
Using $2u_\rho = \alpha'_\rho+\alpha'^*_\rho$ and letting 
\begin{equation}
\alpha_\sigma = \sum\limits_\rho U^*_{\sigma\rho}\,\alpha'_\rho\quad\text{so that}\quad \sum\limits_\sigma\alpha_\sigma\mathbf{a}_\sigma = \sum\limits_\rho \alpha'_\rho\mathbf{a}'_\rho,
\end{equation}
\cref{Arealexpu} is transformed back into \cref{Acplxexp}. Furthermore \cref{dalphaprimedt} becomes
\begin{equation}
\frac{\mathrm{d}\alpha_\sigma}{\mathrm{d}t} = -i\sum\limits_{\sigma'}\left(S_{\sigma\sigma'}\alpha_{\sigma'} + T_{\sigma\sigma'}\alpha^*_{\sigma'}\right)\label{dalphadt}
\end{equation}
where the matrices
\begin{gather}
S = U^*S'U^\text{T}=\frac{D+N^{-1}}{2},\\
D_{\sigma\sigma'} = \frac{1}{\mu_0}\int\mathbf{a}^*_\sigma(\mathbf{r})\cdot\nabla\!\times\!\left(\nabla\!\times\!\mathbf{a}_{\sigma'}(\mathbf{r})\right)\mathrm{d}^3\mathbf{r}\label{Dsigmasigmaprime}\\
\text{and}\quad N_{\sigma\sigma'} = \int\varepsilon(\mathbf{r})\,\mathbf{a}^*_\sigma(\mathbf{r})\cdot\mathbf{a}_{\sigma'}(\mathbf{r})\,\mathrm{d}^3\mathbf{r}\label{Nsigmasigmaprime}
\end{gather}
are all Hermitian, and 
\begin{gather}
T = U^*T'U^\dagger=\frac{\tilde{D}^*-\tilde{N}^{-1}}{2},\\
\tilde{D}_{\sigma\sigma'} = \frac{1}{\mu_0}\int\mathbf{a}_\sigma(\mathbf{r})\cdot\nabla\!\times\!\left(\nabla\!\times\!\mathbf{a}_{\sigma'}(\mathbf{r})\right)\mathrm{d}^3\mathbf{r}\\
\text{and}\quad \tilde{N}_{\sigma\sigma'} = \int\varepsilon(\mathbf{r})\,\mathbf{a}_\sigma(\mathbf{r})\cdot\mathbf{a}_{\sigma'}(\mathbf{r})\,\mathrm{d}^3\mathbf{r}
\end{gather}
are all symmetric.

When working in an orthogonal basis of stationary states, that is one which diagonalises both $N$ and $D$, a useful choice of normalisation for those basis states is to impose the condition $N=D^{-1}$, which makes $T$ vanish (since $T'$ vanishes) and $S=N^{-1}=D$. We can thus say that
\begin{equation}
D_{\sigma\sigma'} = \delta_{\sigma\sigma'}\omega_\sigma\quad\text{and}\quad N_{\sigma\sigma'} = \frac{\delta_{\sigma\sigma'}}{\omega_\sigma}
\end{equation}
where $\omega_\sigma>0$ is the angular frequency of mode $\sigma$ in the sense that $\alpha_\sigma\propto e^{-i\omega_\sigma t}$. In this case, it can be shown that the total electromagnetic energy in the system is
\begin{equation}
E_\text{tot} = 2\sum_\sigma\omega_\sigma|\alpha_\sigma|^2,\label{Etot}
\end{equation}
meaning that $|\alpha_\sigma|^2$ corresponds to $\hbar/2$ times the number of photons in mode $\sigma$. Such a basis, with this normalisation, would always be transformable to a real basis via a block-diagonal unitary matrix in which each block operates within a subspace of states with equal $\omega_\sigma$.

This formalism also works well when $\{\mathbf{a}_\sigma(\mathbf{r})\}$ are not quite stationary states but couple slowly to each other relative to their own natural frequencies, in other words if we can write 
\begin{equation}
D_{\sigma\sigma'} = \delta_{\sigma\sigma'}\bar\omega_\sigma + G_{\sigma\sigma'}\quad\text{and}\quad N_{\sigma\sigma'} = \frac{\delta_{\sigma\sigma'}}{\bar\omega_\sigma} + C_{\sigma\sigma'}\label{perturbedDN}
\end{equation}
where $\bar{\omega}_\sigma$ is the approximate frequency of mode $\sigma$, and for all $\sigma'$, $|G_{\sigma\sigma'}|\ll\bar{\omega}_\sigma$ and $|C_{\sigma\sigma'}|\ll1/\bar\omega_\sigma$. Such a situation could arise if $\{\mathbf{a}_\sigma(\mathbf{r})\}$ are stationary states with eigenfrequencies $\bar\omega_\sigma = \omega_{S,\sigma}$ under a different permittivity profile $\varepsilon_S(\mathbf{r})$ of a subsystem $S$, which is sufficiently similar to $\varepsilon(\mathbf{r})$ that $G_{\sigma\sigma'}$ and $C_{\sigma\sigma'}$ are small. For example, $\varepsilon_S(\mathbf{r})$ could be the permittivity profile of a single waveguide or resonator surrounded everywhere by vacuum, and $\{\mathbf{a}_\sigma(\mathbf{r})\}$ (approximate) stationary states of that subsystem, e.g.~guided modes in the waveguide or whispering gallery modes in the resonator. The overall $\varepsilon(\mathbf{r})$ could describe a system containing more than just that one dielectric body, such that $N_{\sigma\sigma'}$ is only weakly perturbed by the change from $\varepsilon_S(\mathbf{r})$ to $\varepsilon(\mathbf{r})$ associated with introducing the additional bodies. For a number of dielectric bodies coupled to each other in this way, the overall dynamics of guided light can be described in a basis
\begin{equation}
\{\mathbf{a}_\sigma(\mathbf{r})\} = \bigcup\limits_S \left\{\mathbf{a}_{S,\sigma}(\mathbf{r}), \sigma\in\{\sigma\}_S\right\}
\end{equation}
where each $\mathbf{a}_{S,\sigma}(\mathbf{r})$ is a stationary state, with angular frequency $\omega_{S,\sigma}$, of the permittivity profile $\varepsilon_S(\mathbf{r})$. $\{\sigma\}_S$ is the set of values of the label $\sigma$ associated with stationary states of subsystem $S$. If necessary, $\varepsilon_S(\mathbf{r})$ can be modified far from the dielectric in order to keep the modes confined, for instance in the case of whispering gallery modes, which are not true stationary states due to bending losses. Bending and scattering losses can be calculated by including in the basis free travelling wave states of the form $\mathbf{a}_\sigma(\mathbf{r}) = \mathbf{e}_\sigma e^{i\mathbf{k_\sigma}\cdot\mathbf{r}}$, which are stationary states of the vacuum. Calculations of the mode profiles and their coupling strengths for specific geometries are covered elsewhere, particularly in the case of whispering-gallery modes~\cite{chemboModalExpans2010,johnsonTheoryMorph1993,littleAnalyticThe1999}.

Letting $\bar\omega_\sigma = \omega_{S_\sigma,\sigma}$ where $S_\sigma$ denotes the subsystem in which $\mathbf{a}_\sigma(\mathbf{r})$ is a stationary state, i.e.\ $\mathbf{a}_\sigma(\mathbf{r})\in\{\mathbf{a}_{S_\sigma,\sigma}(\mathbf{r})\}$, we will use the normalisation
\begin{equation}
D_{\sigma\sigma'} = \delta_{\sigma\sigma'}\omega_{S_\sigma,\sigma}\quad\text{and}\quad N_{S_\sigma,\sigma\sigma'} = \frac{\delta_{\sigma\sigma'}}{\omega_{S_\sigma,\sigma}}\label{DNSsigma}
\end{equation}
for states $\sigma,\sigma'$ for which $S_{\sigma'} = S_\sigma$, i.e.\ states from the same subsystem, where
\begin{equation}
N_{S_\sigma,\sigma\sigma'} = \int\varepsilon_{S_\sigma}(\mathbf{r})\,\mathbf{a}^*_\sigma(\mathbf{r})\cdot\mathbf{a}_{\sigma'}(\mathbf{r})\,\mathrm{d}^3\mathbf{r}.
\end{equation}
Note that $G_{\sigma\sigma'} = 0$ for states from the same subsystem since $D_{\sigma\sigma'}$ does not depend on $\varepsilon(\mathbf{r})$. For general states $\sigma,\sigma'$ that are not necessarily from the same subsystem, using \cref{perturbedDN,DNSsigma}, we can write
\begin{equation}
G_{\sigma\sigma'} = \omega_{S_\sigma,\sigma}^2C_{S_\sigma,\sigma\sigma'} = \omega_{S_{\sigma'},\sigma'}^2C_{S_{\sigma'},\sigma\sigma'}\label{quickGssp}
\end{equation}
where $C_{S_\sigma,\sigma\sigma'} = N_{S_\sigma,\sigma\sigma'} - \delta_{\sigma\sigma'}/{\omega_{S_\sigma,\sigma}}$. In the limit of small $C_{\sigma\sigma'}$ we have $(N^{-1})_{\sigma\sigma'} = \delta_{\sigma\sigma'}\bar\omega_\sigma - C_{\sigma\sigma'}\bar\omega_\sigma\bar\omega_{\sigma'}$. Although $T$ no longer vanishes, since we are concerned with dynamics on timescales much longer than the inverse optical frequencies, the couplings between $\{\alpha_\sigma\}$ and $\{\alpha^*_\sigma\}$ mediated by $T$ in \cref{dalphadt} can be neglected as they are off-resonant by twice the optical frequency. This means that the dynamics are described by
\begin{equation}
i\frac{\mathrm{d}\alpha_\sigma}{\mathrm{d}t} = \bar\omega_\sigma\alpha_\sigma + \sum\limits_{\sigma'}H_{\sigma\sigma'}\alpha_{\sigma'}\label{Schrodgen}
\end{equation}
where the Hermitian matrix
\begin{equation}
H_{\sigma\sigma'} = \frac{G_{\sigma\sigma'}-C_{\sigma\sigma'}\bar\omega_\sigma\bar\omega_{\sigma'}}{2} = \frac{\bar\omega_\sigma^2 C_{S_\sigma,\sigma\sigma'}-\bar\omega_\sigma\bar\omega_{\sigma'}C_{\sigma\sigma'}}{2}\label{Hsigmasigmaprime}
\end{equation}
can be thought of as the single-photon interaction Hamiltonian divided by $\hbar$. If $|\bar\omega_\sigma - \bar\omega_{\sigma'}| \ll \bar\omega_\sigma$, which must be true in order for the effect of these small coupling terms to be significant, then
\begin{equation}\begin{aligned}
H_{\sigma\sigma'} &= \frac{\bar\omega_\sigma^2 \left(C_{S_\sigma,\sigma\sigma'}-C_{\sigma\sigma'}\right)}{2} = \frac{\bar\omega_\sigma^2 \left(N_{S_\sigma,\sigma\sigma'}-N_{\sigma\sigma'}\right)}{2}\\
&= \frac{\bar\omega_\sigma^2}{2} \int\left(\varepsilon_{S_\sigma}(\mathbf{r}) - \varepsilon(\mathbf{r})\right)\,\mathbf{a}^*_\sigma(\mathbf{r})\cdot\mathbf{a}_{\sigma'}(\mathbf{r})\,\mathrm{d}^3\mathbf{r}.\label{Hsigmasigmaprime2}
\end{aligned}\end{equation}

Losses such as absorption, scattering or bending losses can be included at this point by adding an anti-Hermitian matrix to $H_{\sigma\sigma'}$. Bringing dielectrics together in this way can thus introduce both couplings between confined modes on the same dielectric, leading most notably to frequency splittings between previously degenerate standing-wave modes, and transfer of light between dielectrics. This general approach can also be used in other situations, for example to calculate scattering between free travelling wave states mediated by a dielectric.

\section{Waveguide-resonator coupling}\label{sec:wrc}

Here we are concerned with coupling between guided travelling-wave states in a single-mode tapered optical fiber and whispering-gallery modes in a microresonator. A straight waveguide or sufficiently short section of a tapered optical fiber can be modelled as a permittivity profile $\varepsilon(\mathbf{r}) = \varepsilon(x,y)$. Such a profile will have travelling-wave stationary states $\mathbf{a}_{\tau k}(\mathbf{r}) = \mathbf{a}_{0\tau k}(x,y)\,e^{ikz}$ labelled by their transverse mode index $\tau$ and longitudinal wavevector $k$. The formalism introduced above can be reproduced exactly by assuming that the waveguide has length $L$ with periodic boundary conditions. However, it will then be necessary to let $L\rightarrow\infty$ to simulate an open-ended waveguide with a continuum of $k$ states, which leads to problems with the normalisation of states. We will fix this by replacing instances of $\mathbf{a}_{\tau k}(\mathbf{r})$ and $\alpha_{\tau k}(t)$ with $\mathbf{a}_\tau(k,\mathbf{r})$ and $\alpha_\tau(k,t)$ respectively, defined as follows:
\begin{gather}
\mathbf{a}_\tau(k,\mathbf{r}) = \lim\limits_{L\rightarrow\infty}\sqrt{L}\,\mathbf{a}_{\tau k}(\mathbf{r}) = \mathbf{a}_{0\tau}(k,x,y)\,e^{ikz}\\
\alpha_\tau(k,t) = \lim\limits_{L\rightarrow\infty}\sqrt{L}\,\alpha_{\tau k}(t),
\end{gather}
replacing any sums over $k$ with
\begin{equation}
\lim\limits_{L\rightarrow\infty}\frac{1}{L}\sum\limits_k = \frac{1}{2\pi}\int\mathrm{d}k\label{sumoverktoint}
\end{equation}
and any instances of $\delta_{kk'}$ with $2\pi\delta(k-k')$. Hence we have
\begin{equation}
\mathbf{A}(\mathbf{r},t)=\frac{1}{2\pi}\!\sum\limits_\tau\!\int\!\left(\alpha_\tau(k,t)\mathbf{a}_\tau(k,\mathbf{r})\!+\!\alpha_\tau^*(k,t)\mathbf{a}_\tau^*(k,\mathbf{r})\right)\mathrm{d}k,\label{Artcontinuum}
\end{equation}
\begin{equation}
D_{\tau\tau'}(k,k')=\frac{1}{\mu_0}\int\mathbf{a}^*_\tau(k,\mathbf{r})\cdot\nabla\!\times\!\left(\nabla\!\times\!\mathbf{a}_{\tau'}(k',\mathbf{r})\right)\mathrm{d}^3\mathbf{r}
\end{equation}
and
\begin{align}
N_{\tau\tau'}(k,k') = \int\varepsilon(\mathbf{r})\,\mathbf{a}^*_\tau(k,\mathbf{r})\cdot\mathbf{a}_{\tau'}(k',\mathbf{r})\,\mathrm{d}^3\mathbf{r},
\end{align}
with
\begin{equation}
D_{\tau\tau'}(k,k') = 2\pi\delta(k-k')\,\delta_{\tau\tau'}\,\omega_\tau(k)
\end{equation}
and
\begin{equation}
\quad N_{\tau\tau'}(k,k') = \frac{2\pi\delta(k-k')\,\delta_{\tau\tau'}}{\omega_\tau(k)},
\end{equation}
and thus
\begin{equation}
\frac{1}{\mu_0}\iint\mathbf{a}^*_\tau(k,\mathbf{r})\cdot\nabla\!\times\!\left(\nabla\!\times\!\mathbf{a}_{\tau'}(k',\mathbf{r})\right)\mathrm{d}x\,\mathrm{d}y=\delta_{\tau\tau'}\,\omega_\tau(k)
\end{equation}
and
\begin{equation}
\iint\varepsilon(\mathbf{r})\,\mathbf{a}^*_\tau(k,\mathbf{r})\cdot\mathbf{a}_{\tau'}(k',\mathbf{r})\,\mathrm{d}x\,\mathrm{d}y = \frac{\delta_{\tau\tau'}}{\omega_\tau(k)}.
\end{equation}
\Cref{Etot} becomes
\begin{equation}
E_\text{tot} = \frac{1}{\pi}\sum_\tau\int\omega_\tau(k)|\alpha_\tau(k,t)|^2\mathrm{d}k,\label{Etotaskint}
\end{equation}
meaning that $|\alpha_\tau(k,t)|^2$ is $\pi\hbar$ times the density of photons with respect to $k$. Monochromatic light of wavevector $k_0$ in transverse mode $\tau$ is represented as
\begin{equation}
\alpha_\tau(k,t) = 2\pi A_0\delta(k-k_0)e^{-i\omega_\tau(k_0)t}\label{monochrom}
\end{equation}
which gives
\begin{equation}
\mathbf{A}(\mathbf{r},t) = A_0\,e^{-i\omega_\tau(k_0)}\mathbf{a}_\tau(k_0,\mathbf{r}) + A_0^*\,e^{i\omega_\tau(k_0)}\mathbf{a}^*_\tau(k_0,\mathbf{r}).
\end{equation}
This corresponds to a total electromagnetic energy of $2\omega_\tau(k_0)|A_0|^2$ per unit length, a result that can be derived from~\cref{Etotaskint,monochrom} by substituting one of the factors of $\delta(k-k_0)$ for $\int_{-\infty}^{\infty}e^{i(k-k_0)z}\mathrm{d}z/(2\pi)$. Since for an arbitrarily narrow distribution of wavevectors around $k_0$ the electromagnetic energy travels along the waveguide at the speed of the envelope function, which is the mode's group velocity $v_{\mathrm{g}\,\tau}(k_0)$ defined as
\begin{equation}
v_{\mathrm{g}\,\tau}(k) = \frac{\mathrm{d}\omega_\tau(k)}{\mathrm{d}k},
\end{equation}
the optical power is equal to
\begin{equation}
P = 2\,\omega_\tau(k_0)|A_0|^2\,v_{\mathrm{g}\,\tau}(k_0).
\end{equation}

It is important to note that waveguides and tapered fibers used for coupling light into microresonators are usually single-mode at the operating wavelength, meaning that there are only two possible values of $\tau$, corresponding to the two polarizations of the fundamental transverse mode. Particularly in the case of the fundamental transverse mode, the variation of the transverse mode profile $\mathbf{a}_{0\tau}(k,x,y)$ with $k$ is extremely gradual, taking place over a range of $k$ of the order of $k$ itself, and so can be neglected in the context of a narrow band of optical frequencies. We can thus write $\mathbf{a}_{0\tau}(k,x,y) = \mathbf{a}_{0\tau}(k_0,x,y)$ for a narrow range of $k$ centred around $k_0$. By defining
\begin{equation}
A_\tau(z,t)=\frac{1}{2\pi}\int\alpha_\tau(k,t)e^{i(k-k_0)z}\mathrm{d}k,\label{Atauzt}
\end{equation}
in which the $k$ integral is over this narrow range, we obtain, again in the case where there is only light in transverse mode $\tau$,
\begin{equation}
\mathbf{A}(\mathbf{r},t) = A_\tau(z,t)\,\mathbf{a}_\tau(k_0,\mathbf{r}) +  A^*_\tau(z,t)\,\mathbf{a}^*_\tau(k_0,\mathbf{r}),
\end{equation}
where we can use $\omega_\tau(k) \simeq \omega_\tau(k_0) + v_{\mathrm{g}\,\tau}(k_0)(k-k_0)$ to say that
\begin{equation}
\frac{\partial A_\tau(z,t)}{\partial t} \simeq -i\omega_\tau(k_0) A_\tau(z,t) - v_{\mathrm{g}\,\tau}(k_0) \frac{\partial A_\tau(z,t)}{\partial z}.\label{dAtaudt}
\end{equation}

Bringing a microresonator with whispering gallery modes $\mathbf{a}_\sigma(\mathbf{r})$ close to the waveguide, we may calculate the transfer matrix element $H_{\sigma\,\tau}(k)$ between mode $\mathbf{a}_\sigma(\mathbf{r})$ of the resonator and mode $\mathbf{a}_\tau(k,\mathbf{r})$ of the waveguide using the formula for $H_{\sigma\sigma'}$ given in \cref{Hsigmasigmaprime2} but replacing $\mathbf{a}_{\sigma'}(\mathbf{r})$ with $\mathbf{a}_\tau(k,\mathbf{r})$. Noting that in a system of two dielectrics, $\varepsilon_{S_\sigma}(\mathbf{r})-\varepsilon(\mathbf{r})$ for each body $S_\sigma$ simply equals $-\varepsilon_0$ times the electric susceptibility of the other body, and that $H_{\sigma\sigma'}$ is Hermitian, we obtain
\begin{equation}\begin{aligned}
H_{\sigma\,\tau}(k) &= -\frac{\varepsilon_0\bar\omega^2_\sigma}{2}\int{\chi_\text{wav}(\mathbf{r})\,\mathbf{a}^*_\sigma(\mathbf{r})\cdot\mathbf{a}_\tau(k,\mathbf{r})\,\mathrm{d}^3\mathbf{r}}\\
&= -\frac{\varepsilon_0\bar\omega^2_\sigma}{2}\int{\chi_\text{res}(\mathbf{r})\,\mathbf{a}^*_\sigma(\mathbf{r})\cdot\mathbf{a}_\tau(k,\mathbf{r})\,\mathrm{d}^3\mathbf{r}}
\end{aligned}\end{equation}
where $\chi_\text{wav}(\mathbf{r})$ and $\chi_\text{res}(\mathbf{r})$ are the electric susceptibility profiles of the waveguide and resonator respectively. For $k$ close to $k_0$ as above, we may express this as
\begin{equation}
H_{\sigma\,\tau}(k) = \int\tilde H_{\sigma\,\tau}(k_0,z)\,e^{i(k-k_0)z}\mathrm{d}z\label{HkitoHtildez}
\end{equation}
where
\begin{equation}
\tilde H_{\sigma\,\tau}(k_0,z) \simeq-\frac{\varepsilon_0\bar\omega^2_\sigma}{2}\iint{\chi_\text{res}(\mathbf{r})\,\mathbf{a}^*_\sigma(\mathbf{r})\cdot\mathbf{a}_\tau(k_0,\mathbf{r})\,\mathrm{d}x\,\mathrm{d}y}.
\end{equation}

Thus, if we assume that there is only one resonator mode, namely $\mathbf{a}_\sigma(\mathbf{r})$, that couples significantly to $\mathbf{a}_\tau(k,\mathbf{r})$ for $k$ close to $k_0$ since its frequency is much closer to $\bar\omega_\tau(k_0)$ than that of any other resonator mode, then, combining \cref{Schrodgen} with \cref{dAtaudt} as well as \cref{sumoverktoint,Atauzt,HkitoHtildez}, and adding an intrinsic loss rate $\gamma_0$ to the resonator mode (from processes such as absorption and scattering), we have
\begin{equation}\begin{aligned}
\frac{\partial A_\tau(z,t)}{\partial t}&\simeq-i\bar\omega_\tau(k_0) A_\tau(z,t)\\
 &- v_{\mathrm{g}\,\tau}(k_0) \frac{\partial A_\tau(z,t)}{\partial z} -i\tilde H^*_{\sigma\,\tau}(k_0,z)\,\alpha_\sigma(t)
\end{aligned}\end{equation}
and
\begin{align}
\frac{\mathrm{d}\alpha_\sigma(t)}{\mathrm{d}t}=-(i\bar\omega_\sigma+\gamma_0)\,\alpha_\sigma(t) -i\!\int\!\tilde H_{\sigma\,\tau}(k_0,z)A_\tau(z,t)\,\mathrm{d}z.
\end{align}
Defining the amplitudes $F_\tau(z,t) = A_\tau(z,t)\,e^{i\bar\omega_\tau(k_0)\,t}$ and $\psi_\sigma(t) = \alpha_\sigma(t)\,e^{i\bar\omega_\tau(k_0)\,t}$ in the rotating wave approximation, as well as the detuning $\theta = \bar\omega_\tau(k_0) - \bar\omega_\sigma$, we obtain
\begin{gather}
\frac{\partial F_\tau(z,t)}{\partial t}\simeq - v_{\mathrm{g}\,\tau}(k_0) \frac{\partial F_\tau(z,t)}{\partial z} -i\tilde H^*_{\sigma\,\tau}(k_0,z)\,\psi_\sigma\label{dFtaudt}\\
\frac{\mathrm{d}\psi_\sigma(t)}{\mathrm{d}t}=(i\theta\!-\!\gamma_0)\,\psi_\sigma -i\!\int\!\tilde H_{\sigma\,\tau}(k_0,z)F_\tau(z,t)\,\mathrm{d}z.\label{dpsisigmadt}
\end{gather}
Now for a high-Q resonator, the dynamics of light in a single resonance takes place on a timescale of the inverse cavity linewidth, which is many orders of magnitude larger than the time it takes light to traverse the coupling region (i.e.\ the region where $\tilde H_{\sigma\,\tau}(k_0,z)$ is non-negligible) whilst travelling along the waveguide. Therefore, assuming that the light input into the waveguide is of a linewidth similar to or smaller than the resonance of the cavity (as indeed it must be in order to couple resonantly into it), we may say that $|\partial F_\tau/\partial t| \ll |v_{\mathrm{g}\,\tau}(k_0)\,\partial F_\tau/\partial z|$, allowing us to neglect the left-hand side of \cref{dFtaudt} to obtain
\begin{equation}
\frac{\partial F_\tau(z,t)}{\partial z} = -\frac{i\tilde H^*_{\sigma\,\tau}(k_0,z)}{v_{\mathrm{g}\,\tau}(k_0)}\psi_\sigma.\label{dFtaudz}
\end{equation}
 We thus have $\partial F_\tau(z,t)/\partial z = 0$ outside the coupling region. Defining $F_\text{in}(t)$ and $F_\text{out}(t)$ to be the values of $F_\tau(z,t)$ for $z$ before and after the coupling region respectively, we may integrate \cref{dFtaudz} over $z$ to give
\begin{equation}
F_\text{out}(t) = F_\text{in}(t) - \frac{i H^*_{\sigma\,\tau}(k_0)}{v_{\mathrm{g}\,\tau}(k_0)}\psi_\sigma(t)
\end{equation}
via \cref{HkitoHtildez}. For convenience, we treat integrals over $z$ through the entire coupling region as being between $-\infty$ and $\infty$, meaning that \cref{HkitoHtildez} is equivalent to
\begin{equation}
\tilde H_{\sigma\,\tau}(k_0,z) = \frac{1}{2\pi}\int H_{\sigma\,\tau}(k)\,e^{-i(k-k_0)z}\mathrm{d}k.\label{HtildezitoHk}
\end{equation}
Integrating \cref{dFtaudz} up to an arbitrary $z$ thus gives
\begin{align}
F_\tau(z,t) &= F_\text{in}(t) - \frac{i\psi_\sigma(t)}{v_{\mathrm{g}\,\tau}(k_0)}\int\limits_{-\infty}^z \tilde H^*_{\sigma\,\tau}(k_0,z')\mathrm{d}z'\\
&= F_\text{in}(t) - \frac{i\psi_\sigma(t)}{2\pi v_{\mathrm{g}\,\tau}(k_0)}\int H^*_{\sigma\,\tau}(k)I(k,z)\mathrm{d}k\label{Fztasint}
\end{align}
where
\begin{align}
I(k,z) &= \int\limits_{-\infty}^ze^{i(k-k_0)z'}\mathrm{d}z'\\
&= e^{i(k-k_0)z}\left(\pi\delta(k-k_0)-\frac{i}{k-k_0}\right)\label{Ikzcalcd}.
\end{align}
Substituting for $\tilde H_{\sigma\,\tau}(k_0,z)$ and $F_\tau(z,t)$ in \cref{dpsisigmadt} using \cref{HtildezitoHk} (with the dummy variable $k$ replaced by $k'$) and \cref{Fztasint} respectively, and integrating first over $z$ and then over $k'$, we obtain
\begin{equation}
\frac{\mathrm{d}\psi_\sigma(t)}{\mathrm{d}t} = (i\theta'-\gamma)\,\psi_\sigma(t) - i H_{\sigma\,\tau}(k_0) F_\text{in}(t)\label{dpsisigmadtfinal}
\end{equation}
where $\gamma = \gamma_0+\kappa$, $\theta' = \theta - \delta\omega_\sigma$ and
\begin{gather}
\kappa = \frac{|H_{\sigma\,\tau}(k_0)|^2}{2v_{\mathrm{g}\,\tau}(k_0)}\label{kappaitoHandvg}\\
\delta\omega_\sigma = - \frac{1}{2\pi v_{\mathrm{g}\,\tau}(k_0)}\int \frac{|H_{\sigma\,\tau}(k)|^2}{k-k_0}\,\mathrm{d}k.
\end{gather}
We refer to $\kappa$ as the coupling half-linewidth, to $\gamma_0$ and $\gamma$ as the intrinsic and total half-linewidths respectively, and to $\theta'$ again as the detuning. These expressions can also be derived from Fermi's golden rule and second-order perturbation theory respectively. Although unlikely to be zero, the second-order correction $\delta\omega_\sigma$ to the frequency of the resonator mode will likely be negligible compared to the first-order correction given by $H_{\sigma\sigma}$ that comes from the modification of the permittivity in the vicinity of the resonator due to the waveguide. First-order interaction terms $H_{\tau\tau'}(k,k')$ between the waveguide modes also exist, and have the effect of slightly increasing the wavevector of light as it traverses the coupling region, perhaps in a polarization-dependent way, although this would have little effect on the phenomenology apart from a slight change in the apparent values of the coupling strengths $H_{\sigma\,\tau}(k)$. Bringing the waveguide close to the resonator will also in general increase the effective intrinsic loss rate $\gamma_0$ due to coupling to the other guided mode of the waveguide and to free-space modes.  Note also that momentum-nonconserving couplings between modes in either the waveguide or resonator that are counterpropagating at the coupling region are strongly suppressed due to the fact that the coupling region is uniform over a lengthscale of many wavelengths.

In the steady state where $F_\text{in}$, $F_\text{out}$ and $\psi_\sigma$ are all time-independent, we can thus say that
\begin{equation}
\psi_\sigma = -\frac{i H_{\sigma\,\tau}(k_0) F_\text{in}}{\gamma-i\theta''}\quad\text{and}\quad F_\text{out} = F_\text{in}\left(1-\frac{2\kappa}{\gamma-i\theta''}\right),
\end{equation}
where $\theta''=\theta'-H_{\sigma\sigma}=\bar\omega_\tau(k_0) - \bar\omega_\sigma - \delta\omega_\sigma - H_{\sigma\sigma}$.

The input and output optical powers of the waveguide and stored energy in the cavity are given respectively by
\begin{gather}
P_\text{in,out} = 2\,\bar\omega_\tau(k_0)v_{\mathrm{g}\,\tau}(k_0)|F_\text{in,out}|^2\;\;\text{and}\;\;E_\sigma = 2\bar\omega_\sigma|\psi_\sigma|^2.\label{PinoutandEsigma}
\end{gather}
Thus $E_\sigma$ and $P_\text{out}$ follow Lorentzian profiles with respect to $\theta''$ with half-linewidth $\gamma$, and
\begin{equation}
P_\text{out} = P_\text{in}\left(1-\frac{\eta_\text{in}}{1+(\theta''/\gamma)^2}\right)
\end{equation}
where the in-coupling efficiency $\eta_\text{in} = 4\kappa\gamma_0/\gamma^2$. For a whispering-gallery mode, we may define the circulating power to be
\begin{equation}
P_\text{circ} = E_\sigma\,\Delta\nu_\text{FSR}\label{PcircitoEsigma}
\end{equation}
where $\Delta\nu_\text{FSR}$ is the free spectral range of the mode family in question at mode $\sigma$, which is also the mode's angular group velocity around the resonator divided by $2\pi$.

\section{Kerr nonlinearity}\label{sec:knl}

Turning now to the Kerr effect in the resonator, this adds an extra term $\chi^{(3)}(\mathbf{E}\cdot\mathbf{E})\mathbf{E}$ to the electric polarization vector $\mathbf{P}$ \cite{newIntroduction2011}, where $\mathbf{E}(\mathbf{r}) = -\partial\mathbf{A}/\partial t$ is the electric field. We are assuming a scalar form for $\chi^{(3)}(\mathbf{r})$ as is necessarily true for isotropic materials, and that both the Kerr and linear dielectric effects act intantaneously. If we include this term in Maxwell's equations as part of the displacement field $\mathbf{D} = \varepsilon_0\mathbf{E} + \mathbf{P}$, \cref{AMaxwell} becomes
\begin{equation}
\nabla\!\times\!\left(\nabla\!\times\!\mathbf{A}\right)=\mu_0\!\left(\!\varepsilon\frac{\partial^2\mathbf{A}}{\partial t^2}+\varepsilon_0\chi^{(3)}\frac{\partial}{\partial t}\!\left(\left|\frac{\partial\mathbf{A}}{\partial t}\right|^2\!\frac{\partial\mathbf{A}}{\partial t}\!\right)\!\!\right)\!.\label{AMaxwellKerr} 
\end{equation}
Since this is a small perturbation, we can work in the basis $\{\mathbf{a}_\sigma(\mathbf{r})\}$ of stationary states of \cref{AMaxwell} as previously defined, and let $\alpha_\sigma(t) = \xi_\sigma(t)e^{-i\omega_\sigma t}$ where $|\mathrm{d}\xi_\sigma/\mathrm{d}t|\ll\omega_\sigma|\xi_\sigma|$. To first order in $|\mathrm{d}\xi_\sigma/\mathrm{d}t|/(\omega_\sigma|\xi_\sigma|)$, looking at \cref{Acplxexp}, we have
\begin{equation}\begin{aligned}
\frac{\partial^2\mathbf{A}}{\partial t^2} =& \sum\limits_\sigma\biggl(\!\left(-\omega_\sigma^2\xi_\sigma-2i\omega_\sigma\frac{\mathrm{d}\xi_\sigma}{\mathrm{d}t}\right)e^{-i\omega_\sigma t}\mathbf{a}_\sigma(\mathbf{r})\\
&+\left(-\omega_\sigma^2\xi_\sigma^*+2i\omega_\sigma\frac{\mathrm{d}\xi^*_\sigma}{\mathrm{d}t}\right)e^{i\omega_\sigma t}\mathbf{a}_\sigma^*(\mathbf{r})\biggr).
\end{aligned}\end{equation}
Since the $\chi^{(3)}$ term in \cref{AMaxwellKerr} is already small, we only need to calculate it to leading order, giving
\begin{equation}
-\mu_0\varepsilon_0\chi^{(3)}\frac{\partial}{\partial t}\left(\left(\mathbf{E}_0\cdot\mathbf{E}_0\right)\mathbf{E}_0\right)
\end{equation}
where
\begin{equation}
\mathbf{E}_0(\mathbf{r},t)=i\sum\limits_\sigma\omega_\sigma(\xi_\sigma e^{-i\omega_\sigma t}\mathbf{a}_\sigma(\mathbf{r}) - \xi^*_\sigma e^{i\omega_\sigma t}\mathbf{a}^*_\sigma(\mathbf{r})).\label{E0expansion}
\end{equation}
As the basis states $\{\mathbf{a}_\sigma(\mathbf{r})\}$ are unperturbed, so too is the left-hand side of \cref{AMaxwellKerr} (when expressed in terms of $\{\alpha_\sigma\}$ or $\{\xi_\sigma\}$), so we may equate the total first order perturbation to the right-hand side of \cref{AMaxwellKerr} to zero, which yields
\begin{gather}
2i\varepsilon(\mathbf{r})\sum\limits_\sigma\omega_\sigma\left(\frac{\mathrm{d}\xi_\sigma}{\mathrm{d}t}e^{-i\omega_\sigma t}\mathbf{a}_\sigma(\mathbf{r}) -\frac{\mathrm{d}\xi^*_\sigma}{\mathrm{d}t}e^{i\omega_\sigma t}\mathbf{a}_\sigma^*(\mathbf{r})\right)\nonumber\\
= -\varepsilon_0\chi^{(3)}\frac{\partial}{\partial t}\left(\left(\mathbf{E}_0\cdot\mathbf{E}_0\right)\mathbf{E}_0\right).\label{Kerr1stOrder}
\end{gather}
We may expand the right-hand side as a triple sum over $\rho$, $\mu$ and $\nu$ by expressing each instance of $\mathbf{E}_0$ in the form given in \cref{E0expansion}, but with the index $\sigma$ replaced by $\rho$, $\mu$ and $\nu$ respectively. Doing this, we see that for one of the resulting eight terms to be resonant with the positive-frequency ($e^{-i\omega_\sigma t}$) term on the left-hand side it must satisfy $\omega_\sigma\pm\omega_\rho\pm\omega_\mu\pm\omega_\nu\simeq0$ for some combination of plus and minus signs. Terms that satisfy this with one or three minus signs correspond to processes that convert one photon into three or vice versa, such as third-harmonic generation, and will not be discussed here. We are interested in terms that satisfy it with two minus signs, that correspond to processes that conserve the total photon number and can thus operate entirely within a single narrow band of optical frequencies. As explained below, these processes comprise self- and cross-phase modulation (which cause frequency shifts of modes) and four-wave mixing (which transfers light between modes), although in a given situation the choice of term may depend on the mode basis being used. Thus, taking the dot product of both sides of \cref{Kerr1stOrder} with $\mathbf{a}^*_\sigma(\mathbf{r})$ and integrating over all space, noting the normalisation $N_{\sigma\sigma'} = \delta_{\sigma\sigma'}/\omega_\sigma$, equating the $e^{-i\omega_\sigma t}$ terms on each side and using the fact that to leading order the $\mathrm{d}/\mathrm{d}t$ on the right-hand side simply multiplies these by $-i\omega_\sigma$, we obtain
\begin{equation}
\frac{\mathrm{d}\xi_\sigma}{\mathrm{d}t} = i\sum\limits_\rho\sum\limits_\mu\sum\limits_\nu K_{\sigma\rho\mu\nu}\,\xi^*_\rho\xi_\mu\xi_\nu e^{i(\omega_\sigma+\omega_\rho-\omega_\mu-\omega_\nu)t}
\end{equation}
or equivalently
\begin{equation}
\frac{\mathrm{d}\alpha_\sigma}{\mathrm{d}t} = -i\omega_\sigma\alpha_\sigma +i\sum\limits_\rho\sum\limits_\mu\sum\limits_\nu K_{\sigma\rho\mu\nu}\,\alpha^*_\rho\alpha_\mu\alpha_\nu
\end{equation}
where
\begin{equation}\begin{aligned}
K_{\sigma\rho\mu\nu} &= \frac{\varepsilon_0}{2}\omega_\sigma\omega_\rho\omega_\mu\omega_\nu\!\int\!\chi^{(3)}\bigl((\mathbf{a}^*_\sigma\!\cdot\!\mathbf{a}^*_\rho)(\mathbf{a}_\mu\!\cdot\!\mathbf{a}_\nu)\\
&+(\mathbf{a}^*_\sigma\!\cdot\!\mathbf{a}_\mu)(\mathbf{a}^*_\rho\!\cdot\!\mathbf{a}_\nu)+(\mathbf{a}^*_\sigma\!\cdot\!\mathbf{a}_\nu)(\mathbf{a}^*_\rho\!\cdot\!\mathbf{a}_\mu)\bigr)\mathrm{d}^3\mathbf{r}.\label{Ksrmn}
\end{aligned}\end{equation}
Terms with $\sigma=\rho=\mu=\nu$ correspond to self-phase modulation (SPM), which can be seen as coming from a change in the refractive index seen by a light wave that is proportional to the wave's own local intensity. For a linearly polarised travelling-wave mode,
\begin{equation}
K_{\sigma\sigma\sigma\sigma} = \frac{3\varepsilon_0\omega_\sigma^4}{2}\int\chi^{(3)}\|\mathbf{a}_\sigma\|^4\mathrm{d}^3\mathbf{r}.
\end{equation}
Observing that this term results in a self-induced frequency shift $\Delta\omega_{\sigma\sigma} = -K_{\sigma\sigma\sigma\sigma}|\alpha_\sigma|^2$, we may use this to calculate the change in refractive index for a given optical intensity by treating a plane wave in an infinite uniform medium as though it is propagating inside a cuboid with volume $V$ and periodic boundary conditions. We equate $\Delta\omega_{\sigma\sigma}/\omega_\sigma$ to $-\Delta n/n_0$ where $\Delta n$ is this change in refractive index and $n_0 = \sqrt{\varepsilon/\varepsilon_0}$ is the linear refractive index. Noting that the optical intensity is $I=2\omega_\sigma|\alpha_\sigma|^2c/(n_0V)$ where $c$ is the speed of light in a vacuum and that 
\begin{equation}
K_{\sigma\sigma\sigma\sigma}=\frac{3\omega_\sigma^2\chi^{(3)}}{2n_0^2\varepsilon V}\label{Ksssssimple}
\end{equation}
given the normalisation of $\mathbf{a}_\sigma$, we can show that
\begin{equation}
\Delta n = n_2 I,\quad\text{where}\quad n_2 = \frac{3\chi^{(3)}}{4\varepsilon c}
\end{equation}
is known as the nonlinear refractive index. We can generalise \cref{Ksssssimple} to any optical mode in a resonator by defining the effective mode volume to be
\begin{equation}
V_\sigma = \frac{1}{\omega_\sigma^2\varepsilon_\text{res}^2\int_\text{res}{\|\mathbf{a}_\sigma(\mathbf{r})\|^4\mathrm{d}^3\mathbf{r}}} = \frac{\left(\int{\varepsilon(\mathbf{r})\|\mathbf{a}_\sigma(\mathbf{r})\|^2\mathrm{d}^3\mathbf{r}}\right)^2}{\varepsilon_\text{res}^2\int_\text{res}{\|\mathbf{a}_\sigma(\mathbf{r})\|^4\mathrm{d}^3\mathbf{r}}}
\end{equation}
where $\varepsilon_\text{res}$ is the value of $\varepsilon$ in the resonator and subscript ``res'' on the bottom integral indicates that it is only over the volume of the resonator itself, as opposed to the top integral which is over all space including any evanescent regions outside the resonator. We have assumed that $\chi^{(3)}$ is a constant inside the resonator and zero outside it, as is the case for any resonator made of a homogeneous material. Thus
\begin{equation}
K_{\sigma\sigma\sigma\sigma}=\frac{2cn_2\omega_\sigma^2}{n_0^2 V_\sigma}\label{Kssssiton2}
\end{equation}
where $n_2$ and $n_0$ refer to their values inside the resonator.

Each mode also experiences frequency shifts proportional to the intensities of light in the other modes, due to terms in which $\sigma = \mu$ and $\rho = \nu$, or $\sigma = \nu$ and $\rho = \mu$, but $\sigma\ne\rho$. Known as cross-phase modulation (XPM), the value of this shift induced on mode $\sigma$ by mode $\rho$ is thus given by $\Delta\omega_{\sigma\rho} = -2K_{\sigma\rho\sigma\rho}|\alpha_\rho|^2$, since $K_{\sigma\rho\mu\nu} = K_{\sigma\rho\nu\mu}$. All other terms transfer light between modes, and are collectively known as four-wave mixing. Importantly, in systems with a high degree of symmetry such as a WGM resonator with rotational symmetry, most of the terms of $K_{\sigma\rho\mu\nu}$ will turn out to be zero. These cases can be understood by realising that quantum-mechanically the $K_{\sigma\rho\mu\nu}$ term is annihilating a photon in each of modes $\mu$ and $\nu$ and creating one in each of modes $\sigma$ and $\rho$, and must conserve the total linear or angular momentum in the cases of translational and rotational symmetry respectively. Thus for whispering-gallery modes, in order to conserve angular momentum, the sum of the azimuthal mode numbers of modes $\sigma$ and $\rho$ must equal that of modes $\mu$ and $\nu$ in order for $K_{\sigma\rho\mu\nu}$ to be non-zero. In WGM resonators, distinct modes with the same azimuthal mode number tend to differ in frequency by more than the free spectral range of the resonator. This is due to the strong radial and axial confinement that splits different radially- and axially-excited modes, as well as to the strong geometric birefringence that splits the radially- and axially-polarized versions of the same spatial mode. As a consequence, terms of the form $K_{\sigma\rho\sigma\mu}$ or $K_{\sigma\rho\mu\sigma}$ with $\rho\ne\mu$ will usually be strongly off-resonant and thus negligible. Therefore the total Kerr frequency shift of mode $\sigma$ contains only the SPM and XPM terms already discussed, and so is given by
\begin{equation}
\Delta\omega_\sigma = -K_{\sigma\sigma\sigma\sigma}|\alpha_\sigma|^2 -2\sum\limits_{\rho\ne\sigma}K_{\sigma\rho\sigma\rho}|\alpha_\rho|^2.\label{Deltaomegasigma}
\end{equation}
By examining terms of the form $K_{\sigma\rho\sigma\rho}$ $(=K_{\sigma\rho\rho\sigma})$ and $K_{\sigma\sigma\rho\rho}$ in \cref{Ksrmn} and applying symmetry considerations, we can derive some important results about the relative magnitudes of these SPM and XPM shifts in various cases. We start by noting that $K_{\sigma\rho\sigma\rho}$ is invariant under multiplication of either $\mathbf{a}_\sigma(\mathbf{r})$ or $\mathbf{a}_\rho(\mathbf{r})$ by a spatially dependent phase factor $e^{i\varphi(\mathbf{r})}$. Now travelling-wave modes in a resonator (waveguide) belong to mode families, which are series (continua) of modes that differ only by their azimuthal mode number (longitudinal wavevector). Modes from the same family, particularly those that are close in this mode number or wavevector, have essentially the same spatial mode profile up to a multiplicative spatially dependent phase factor. They will therefore have $K_{\sigma\rho\sigma\rho} \simeq K_{\sigma\sigma\sigma\sigma} \simeq K_{\rho\rho\rho\rho}$ and hence XPM that is almost exactly twice as strong as SPM (by \cref{Deltaomegasigma}). Furthermore, all travelling-wave modes have a counterpropagating but otherwise identical counterpart, whose mode profile $\mathbf{a}_\sigma(\mathbf{r})$ is the complex conjugate of that of the first mode. This can be seen by from \cref{Acplxexp} by noticing that if $\alpha_\sigma(t) \propto e^{-i\omega_\sigma t}$ then swapping $\mathbf{a}_\sigma(\mathbf{r})$ and $\mathbf{a}_\sigma^*(\mathbf{r})$ is equivalent to exchanging $t$ and $-t$. A perfect travelling-wave mode is one that has a distinct counterpropagating counterpart, in other words if $\mathbf{a}_\sigma(\mathbf{r})$ and $\mathbf{a}_{\sigma'}(\mathbf{r})=\mathbf{a}_\sigma^*(\mathbf{r})$ are orthogonal as defined by the matrix elements $N_{\sigma\sigma'}$ (and $D_{\sigma\sigma'}$) between them vanishing. The antithesis of a travelling-wave mode is a standing-wave mode, for which $\mathbf{a}_\sigma^*(\mathbf{r}) = \mathbf{a}_\sigma(\mathbf{r})$ as stated in \cref{sec:rmc}. It can be seen from \cref{Ksrmn} that if $\mathbf{a}_\mu(\mathbf{r}) = \mathbf{a}^*_\rho(\mathbf{r})$ then $K_{\sigma\mu\sigma\mu} = K_{\sigma\rho\sigma\rho}$, meaning that in a travelling-wave basis the strength of XPM between any two modes is exactly the same as between the first mode and the counterpropagating partner of the second. Crucially for this paper, it also implies that XPM is precisely twice as strong as SPM for modes that are counterpropagating partners of each other.

Finding the XPM-SPM ratio between modes of different polarizations is a little more nuanced. In the case of plane waves or in the limit of weakly guided waves, $\mathbf{a}_\mu(\mathbf{r})$ is everywhere perpendicular to the propagation direction, and every mode has a counterpart with a different polarization but otherwise the same spatial mode profile. This can be seen from the fact that in isotropic media with only small variations in refractive index, Maxwell's equations can be approximated by the same scalar wave equation for both polarizations~\cite{buresGuidedOptic2009}. For example, a spatial mode propagating along $\mathbf{e}_z$ will be have two linearly polarized modes that can be expressed as $\mathbf{a}_x(\mathbf{r})=\mathbf{e}_x a(\mathbf{r})$ and $\mathbf{a}_y(\mathbf{r})=\mathbf{e}_y a(\mathbf{r})$ for some common scalar function $a(\mathbf{r})$, where $\mathbf{e}_{x,y,z}$ are the Cartesian unit vectors. \cref{Ksrmn} thus tells us that 
\begin{gather}
K_{xxxx}=K_{yyyy}=3K_0\\
\text{and}\quad K_{xyxy}=K_{xxyy}=K_{yyxx}=K_0,
\end{gather}
where
\begin{equation}
K_0=\frac{\varepsilon_0}{2}\omega^4\int\chi^{(3)}(\mathbf{r})\,|a(\mathbf{r})|^4\,\mathrm{d}^3\mathbf{r}
\end{equation}
and $\omega=\omega_x=\omega_y$. This means that for oppositely linearly polarized counterparts, XPM is $2/3$ as strong as SPM. However, light is also transferred between the two polarizations due to the $K_{xxyy}$ and $K_{yyxx}$ terms, and so a linearly polarized basis is only appropriate for describing the physics if there is sufficient mode splitting in that basis to suppress this transfer (such as in a WGM resonator or rectangular waveguide, both of which have strong geometric berefringence). In the absence of such a mode splitting, a better basis to use is a circularly polarized one consisting of $\mathbf{a}_+(\mathbf{r})=\mathbf{e}_+ a(\mathbf{r})$ and $\mathbf{a}_-(\mathbf{r})=\mathbf{e}_- a(\mathbf{r})$, where $\mathbf{e}_\pm=(\mathbf{e}_x\pm i\mathbf{e}_y)/\sqrt{2}$. In this basis, \cref{Ksrmn} yields
\begin{gather}
K_{++++}=K_{----}=K_{+-+-}=2K_0\label{Kppppeq2K0}\\
\text{and}\quad K_{++--}=K_{--++}=0\label{Kppmmeq0}.
\end{gather}
Thus, SPM is only $2/3$ as strong as it is in a linearly polarized basis, however XPM is now twice as strong as SPM and there is no longer any transfer between the two modes. \cref{Kppmmeq0} can be justified via conservation of spin angular momentum, since if it were not true, two photons with $+1$ spin angular momentum would be able to convert into two with $-1$ and vice versa. Furthermore, the equality of $K_{+-+-}$ with $K_{++++}$ and $K_{----}$ can be deduced from the fact that $\mathbf{a}_+(\mathbf{r})$ is equal to $\mathbf{a}_-^*(\mathbf{r})$ multiplied by some spatially dependent phase factor $e^{i\varphi(\mathbf{r})}$, as explained earlier. For strongly guided modes, i.e.\ ones with transverse features not much larger than the wavelength, this no longer holds due to the significant component of $\mathbf{a}_+(\mathbf{r})$ pointing along the propagation direction, meaning that XPM between oppositely circularly polarized modes is less than twice as strong as SPM.

Turning again to four-wave mixing, in cases where $\alpha_\sigma$ and $\alpha_\rho$ are initially both zero, the process governed by $K_{\sigma\rho\mu\nu}$ will only occur when $|\alpha_\mu\alpha_\nu|$ surpasses a certain threshold where the gain in $\alpha_\sigma$ and $\alpha_\rho$ through mutual positive feedback becomes greater than their losses. This is true for sideband and frequency comb generation starting from monochromatic light. Since this is also governed by the Kerr effect, its threshold power is roughly the same as that for the symmetry breaking effect between counterpropagating light mentioned in \cref{sec:intro} and studied in Refs.~\cite{caoExperimentalDemonstration2017,delbinoSymmetryBreaking2017,woodleyUniversalSymmetrybreaking2018,silverCriticalDyn2019}, and in fact is normally higher due to dispersion in the resonator. Therefore it is usually possible to pump a pair of counterpropagating modes with sufficient power to observe the symmetry breaking but no other Kerr nonlinear processes.

Thus, returning to \cref{dpsisigmadtfinal}, letting $\sigma=1,2$ denote two counterpropagating partner modes along with waveguide input field amplitudes $F_\text{in,1,2}(t)$ in the corresponding directions and including the SPM and XPM frequency shifts, we obtain 
\begin{align}
\frac{\mathrm{d}\psi_{1,2}}{\mathrm{d}t}=&\left(i\theta''_{1,2}+iK\left(|\psi_{1,2}|^2+2|\psi_{2,1}|^2\right)-\gamma\right)\psi_{1,2}\nonumber\\
&-iHF_\text{in,1,2},
\end{align}
where $\theta''_{1,2}$ are the detunings of the pumps in each direction from the resonance without Kerr shift, $H$ denotes the value of $H_{\sigma\,\tau}(k_0)$ between each resonator mode and the copropagating waveguide mode, and $K=K_{1111}=K_{2222}=K_{1212}=K_{2121}$. The values of $H_{\sigma\,\tau}(k_0)$ for each direction are the same by symmetry, with any difference due to a difference in pump frequency being negligible, and linear couplings between counterpropagating modes are assumed to be negligible. Finally, we may put this in dimensionless form by letting
\begin{equation}\begin{gathered}
\bar t = \gamma t,\quad \Delta_{1,2} = -\frac{\theta''_{1,2}}{\gamma},\quad e_{1,2} = \sqrt{\frac{K}{\gamma}}\psi^*_{1,2},\\
\tilde e_{1,2} = iH^*\sqrt{\frac{K}{\gamma^3}}F^*_\text{in,1,2},\quad \dot e_{1,2} = \frac{\mathrm{d}e_{1,2}}{\mathrm{d}\bar t},
\end{gathered}\end{equation}
yielding
\begin{equation}
\dot{e}_{1,2} = \tilde{e}_{1,2} - \left(1+i\left( |e_{1,2}|^2 + 2|e_{2,1}|^2 - \Delta_{1,2}\right)\right)e_{1,2},\label{dynEq}
\end{equation}
which forms the basis of the analysis of the symmetry-breaking dynamics in Refs.~\cite{woodleyUniversalSymmetrybreaking2018,silverCriticalDyn2019}. \cref{tab:dlqs} provides a more empirical set of definitions for the quantities in~\cref{dynEq} that mirror those in Refs.~\cite{woodleyUniversalSymmetrybreaking2018,silverCriticalDyn2019}, in which $|\tilde{e}_{1,2}|^2$ and $|e_{1,2}|^2$ are the dimensionless pump and circulating powers $\tilde{p}_{1,2}$ and $p_{1,2}$ respectively. These definitions may be reconciled with the rest of this paper by examining \cref{kappaitoHandvg,PinoutandEsigma,PcircitoEsigma,Kssssiton2}, substituting $\bar\omega_\tau(k_0)$, $\bar\omega_\sigma$ and $\omega_\sigma$ with $\omega_0$ and $V_\sigma$ with $V$.

\begin{table}
\caption{Definition of dimensionless quantities in~\cref{dynEq}. $\eta_\text{in}$ is the resonant in-coupling efficiency equal to $4\kappa\gamma_0/\gamma^2$ where $\kappa$, $\gamma_0$ and $\gamma=\gamma_0+\kappa$ are the coupling, intrinsic and total half-linewidths respectively. $P_{\text{in,}1,2}$ and $P_{\text{circ,}1,2}$ are the pump and circulating powers respectively. $P_0 = \pi n_0^2V/(n_2\lambda QQ_0)$ is the characteristic in-coupled power required for Kerr nonlinear effects, where $n_0$ and $n_2$ are the linear and nonlinear refractive indices, $V$ is the mode volume, and $Q=\omega_0/(2\gamma)$ and $Q_0=\omega_0/(2\gamma_0)$ are the loaded and intrinsic quality factors respectively for cavity resonance frequency $\omega_0$ (without Kerr shift). $\mathcal{F}_0 = \Delta\omega_\text{FSR}/(2\gamma_0)$ is the cavity's intrinsic finesse for free spectral range $\Delta\omega_\text{FSR}$, and $\omega_{1,2}$ are the pump frequencies.}
\label{tab:dlqs}
  \begin{center}
    \begin{tabular}{clc}
      \toprule
      \textbf{Symbol} & \textbf{Description} & \textbf{Formula}\\
      \colrule
      $\tilde{p}_{1,2}$ & Pump powers & $\eta_\text{in}P_{\text{in,}1,2}/P_0$\\
      $p_{1,2}$ & Circulating powers & $2\pi P_{\text{circ,}1,2}/(\mathcal{F}_0 P_0)$\\
      $\Delta_{1,2}$ & \parbox{4cm}{\raggedright \vspace{1.2mm} Pump detunings from resonance frequency without Kerr shift\strut} & $(\omega_0-\omega_{1,2})/\gamma$\\
      $\tilde{e}_{1,2}$ & Pump field amplitudes & $\tilde{p}_{1,2} = \left|\tilde{e}_{1,2}\right|^2$\\
      $e_{1,2}$ & Circulating field amplitudes & $p_{1,2} = \left|e_{1,2}\right|^2$\\
      \botrule
    \end{tabular}
  \end{center}
\end{table}

\section{Conclusion}

We have brought together the various elements of the coupled mode theory descriptions of linear coupling and Kerr interaction between modes of a dielectric optical microresonator and a waveguide, starting from first principles. The treatment is initially very general and not specific to a particular geometry or choice of mode basis, and can thus be applied to many scenarios not discussed here such as geometric scattering between resonator modes, bending losses and losses due to surface roughness. We then used this theory to derive the dimensionless equation governing the symmetry-breaking dynamics of a pair of counterpropagating modes in a WGM or ring resonator, proving that the factor of two between the coefficients of SPM and XPM is exact when the two modes are time-reversal conjugates of each other. This factor is slightly less than two for modes of opposite circular polarization and/or different frequency, due to small differences between the two spatial mode profiles. All the approximations used in this paper are essentially based on the same assumption, that all the dynamical processes in the resonator (decay of light, coupling of light from and to the waveguide, and Kerr interaction) occur on timescales much longer than the inverse optical frequency. They are therefore valid to a very high degree of accuracy on the order of $1/Q$, where the quality factor $Q$ of the resonator is generally at least $10^6$ (sometimes even exceeding $10^{10}$) for resonators used to realise Kerr nonlinear effects~\cite{kippenbeMicroresonat2011}. The method and assumptions used to describe a continuum of optical modes of a translationally symmetric waveguide in terms of a complex field variable of a single spatial dimension can be easily adapted to describe a mode family of a rotationally symmetric WGM resonator, allowing the LLE to be derived from the terms already discussed plus one or more dispersion terms.

\section{Acknowledgments}

This work was supported by the Royal Academy of Engineering and the Office of the Chief Science Adviser for National Security under the UK Intelligence Community Postdoctoral Fellowship Programme. P.\ D.\ H.\ acknowledges funding from Horizon 2020 European Research Council (ERC) (No. 756966, CounterLight).


\begin{thebibliography}{61}%
\makeatletter
\providecommand \@ifxundefined [1]{%
 \@ifx{#1\undefined}
}%
\providecommand \@ifnum [1]{%
 \ifnum #1\expandafter \@firstoftwo
 \else \expandafter \@secondoftwo
 \fi
}%
\providecommand \@ifx [1]{%
 \ifx #1\expandafter \@firstoftwo
 \else \expandafter \@secondoftwo
 \fi
}%
\providecommand \natexlab [1]{#1}%
\providecommand \enquote  [1]{``#1''}%
\providecommand \bibnamefont  [1]{#1}%
\providecommand \bibfnamefont [1]{#1}%
\providecommand \citenamefont [1]{#1}%
\providecommand \href@noop [0]{\@secondoftwo}%
\providecommand \href [0]{\begingroup \@sanitize@url \@href}%
\providecommand \@href[1]{\@@startlink{#1}\@@href}%
\providecommand \@@href[1]{\endgroup#1\@@endlink}%
\providecommand \@sanitize@url [0]{\catcode `\\12\catcode `\$12\catcode
  `\&12\catcode `\#12\catcode `\^12\catcode `\_12\catcode `\%12\relax}%
\providecommand \@@startlink[1]{}%
\providecommand \@@endlink[0]{}%
\providecommand \url  [0]{\begingroup\@sanitize@url \@url }%
\providecommand \@url [1]{\endgroup\@href {#1}{\urlprefix }}%
\providecommand \urlprefix  [0]{URL }%
\providecommand \Eprint [0]{\href }%
\providecommand \doibase [0]{http://dx.doi.org/}%
\providecommand \selectlanguage [0]{\@gobble}%
\providecommand \bibinfo  [0]{\@secondoftwo}%
\providecommand \bibfield  [0]{\@secondoftwo}%
\providecommand \translation [1]{[#1]}%
\providecommand \BibitemOpen [0]{}%
\providecommand \bibitemStop [0]{}%
\providecommand \bibitemNoStop [0]{.\EOS\space}%
\providecommand \EOS [0]{\spacefactor3000\relax}%
\providecommand \BibitemShut  [1]{\csname bibitem#1\endcsname}%
\let\auto@bib@innerbib\@empty
%</preamble>
\bibitem [{\citenamefont {Braginsky}\ \emph {et~al.}(1989)\citenamefont
  {Braginsky}, \citenamefont {Gorodetsky},\ and\ \citenamefont {{V. S.
  Ilchenko}}}]{braginskQualityFacto1989}%
  \BibitemOpen
  \bibfield  {author} {\bibinfo {author} {\bibfnamefont {V.~B.}\ \bibnamefont
  {Braginsky}}, \bibinfo {author} {\bibfnamefont {M.~L.}\ \bibnamefont
  {Gorodetsky}}, \ and\ \bibinfo {author} {\bibnamefont {{V. S. Ilchenko}}},\
  }\href {\doibase 10.1016/0375-9601(89)90912-2} {\bibfield  {journal}
  {\bibinfo  {journal} {Physics Letters A}\ }\textbf {\bibinfo {volume}
  {137}},\ \bibinfo {pages} {393} (\bibinfo {year} {1989})}\BibitemShut
  {NoStop}%
\bibitem [{\citenamefont {Vahala}(2003)}]{vahalaOpticalMicr2003}%
  \BibitemOpen
  \bibfield  {author} {\bibinfo {author} {\bibfnamefont {K.~J.}\ \bibnamefont
  {Vahala}},\ }\href {\doibase 10.1038/nature01939} {\bibfield  {journal}
  {\bibinfo  {journal} {Nature}\ }\textbf {\bibinfo {volume} {424}},\ \bibinfo
  {pages} {839} (\bibinfo {year} {2003})}\BibitemShut {NoStop}%
\bibitem [{\citenamefont {Johnson}(1993)}]{johnsonTheoryMorph1993}%
  \BibitemOpen
  \bibfield  {author} {\bibinfo {author} {\bibfnamefont {B.~R.}\ \bibnamefont
  {Johnson}},\ }\href {\doibase 10.1364/JOSAA.10.000343} {\bibfield  {journal}
  {\bibinfo  {journal} {Journal of the Optical Society of America A-optics
  Image Science and Vision}\ }\textbf {\bibinfo {volume} {10}},\ \bibinfo
  {pages} {343} (\bibinfo {year} {1993})}\BibitemShut {NoStop}%
\bibitem [{\citenamefont {Marcuse}(1985)}]{marcuseCoupledMode1985}%
  \BibitemOpen
  \bibfield  {author} {\bibinfo {author} {\bibfnamefont {D.}~\bibnamefont
  {Marcuse}},\ }\href {\doibase 10.1109/JQE.1985.1072590} {\bibfield  {journal}
  {\bibinfo  {journal} {IEEE J. Quantum Electron.}\ }\textbf {\bibinfo {volume}
  {21}},\ \bibinfo {pages} {1819} (\bibinfo {year} {1985})}\BibitemShut
  {NoStop}%
\bibitem [{\citenamefont {Stoffer}\ \emph {et~al.}(2005)\citenamefont
  {Stoffer}, \citenamefont {Hiremath}, \citenamefont {Hammer}, \citenamefont
  {Prkna},\ and\ \citenamefont {{\v
  C}tyrok{\'y}}}]{stofferCylindricalIntegrated2005}%
  \BibitemOpen
  \bibfield  {author} {\bibinfo {author} {\bibfnamefont {R.}~\bibnamefont
  {Stoffer}}, \bibinfo {author} {\bibfnamefont {K.~R.}\ \bibnamefont
  {Hiremath}}, \bibinfo {author} {\bibfnamefont {M.}~\bibnamefont {Hammer}},
  \bibinfo {author} {\bibfnamefont {L.}~\bibnamefont {Prkna}}, \ and\ \bibinfo
  {author} {\bibfnamefont {J.}~\bibnamefont {{\v C}tyrok{\'y}}},\ }\href
  {\doibase 10.1016/j.optcom.2005.06.064} {\bibfield  {journal} {\bibinfo
  {journal} {Optics Communications}\ }\textbf {\bibinfo {volume} {256}},\
  \bibinfo {pages} {46} (\bibinfo {year} {2005})}\BibitemShut {NoStop}%
\bibitem [{\citenamefont {Schelkunoff}(1955)}]{schelkunConversionM1955}%
  \BibitemOpen
  \bibfield  {author} {\bibinfo {author} {\bibfnamefont {S.~A.}\ \bibnamefont
  {Schelkunoff}},\ }\href {\doibase 10.1002/j.1538-7305.1955.tb03787.x}
  {\bibfield  {journal} {\bibinfo  {journal} {The Bell System Technical
  Journal}\ }\textbf {\bibinfo {volume} {34}},\ \bibinfo {pages} {995}
  (\bibinfo {year} {1955})}\BibitemShut {NoStop}%
\bibitem [{\citenamefont {Yariv}(1973)}]{yarivCoupledmode1973}%
  \BibitemOpen
  \bibfield  {author} {\bibinfo {author} {\bibfnamefont {A.}~\bibnamefont
  {Yariv}},\ }\href {\doibase 10.1109/JQE.1973.1077767} {\bibfield  {journal}
  {\bibinfo  {journal} {IEEE Journal of Quantum Electronics}\ }\textbf
  {\bibinfo {volume} {9}},\ \bibinfo {pages} {919} (\bibinfo {year}
  {1973})}\BibitemShut {NoStop}%
\bibitem [{\citenamefont {Haus}\ \emph {et~al.}(1987)\citenamefont {Haus},
  \citenamefont {Huang}, \citenamefont {Kawakami},\ and\ \citenamefont
  {Whitaker}}]{hausCoupledmode1987}%
  \BibitemOpen
  \bibfield  {author} {\bibinfo {author} {\bibfnamefont {H.}~\bibnamefont
  {Haus}}, \bibinfo {author} {\bibfnamefont {W.}~\bibnamefont {Huang}},
  \bibinfo {author} {\bibfnamefont {S.}~\bibnamefont {Kawakami}}, \ and\
  \bibinfo {author} {\bibfnamefont {N.}~\bibnamefont {Whitaker}},\ }\href
  {\doibase 10.1109/JLT.1987.1075416} {\bibfield  {journal} {\bibinfo
  {journal} {J. Lightwave Technol.}\ }\textbf {\bibinfo {volume} {5}},\
  \bibinfo {pages} {16} (\bibinfo {year} {1987})}\BibitemShut {NoStop}%
\bibitem [{\citenamefont {Haus}\ and\ \citenamefont
  {Huang}(1991)}]{hausCoupledmode1991}%
  \BibitemOpen
  \bibfield  {author} {\bibinfo {author} {\bibfnamefont {H.}~\bibnamefont
  {Haus}}\ and\ \bibinfo {author} {\bibfnamefont {W.}~\bibnamefont {Huang}},\
  }\href {\doibase 10.1109/5.104225} {\bibfield  {journal} {\bibinfo  {journal}
  {Proceedings of the IEEE}\ }\textbf {\bibinfo {volume} {79}},\ \bibinfo
  {pages} {1505} (\bibinfo {year} {1991})}\BibitemShut {NoStop}%
\bibitem [{\citenamefont {Little}\ \emph {et~al.}(1999)\citenamefont {Little},
  \citenamefont {Laine},\ and\ \citenamefont {Haus}}]{littleAnalyticThe1999}%
  \BibitemOpen
  \bibfield  {author} {\bibinfo {author} {\bibfnamefont {B.~E.}\ \bibnamefont
  {Little}}, \bibinfo {author} {\bibfnamefont {J.~P.}\ \bibnamefont {Laine}}, \
  and\ \bibinfo {author} {\bibfnamefont {H.~A.}\ \bibnamefont {Haus}},\
  }\href@noop {} {\bibfield  {journal} {\bibinfo  {journal} {Journal Of
  Lightwave Technology}\ }\textbf {\bibinfo {volume} {17}},\ \bibinfo {pages}
  {704} (\bibinfo {year} {1999})}\BibitemShut {NoStop}%
\bibitem [{\citenamefont {Spillane}\ \emph {et~al.}(2003)\citenamefont
  {Spillane}, \citenamefont {Kippenberg}, \citenamefont {Painter},\ and\
  \citenamefont {Vahala}}]{spillaneIdealityFib2003}%
  \BibitemOpen
  \bibfield  {author} {\bibinfo {author} {\bibfnamefont {S.~M.}\ \bibnamefont
  {Spillane}}, \bibinfo {author} {\bibfnamefont {T.~J.}\ \bibnamefont
  {Kippenberg}}, \bibinfo {author} {\bibfnamefont {O.~J.}\ \bibnamefont
  {Painter}}, \ and\ \bibinfo {author} {\bibfnamefont {K.~J.}\ \bibnamefont
  {Vahala}},\ }\href {\doibase 10.1103/PhysRevLett.91.043902} {\bibfield
  {journal} {\bibinfo  {journal} {Phys. Rev. Lett.}\ }\textbf {\bibinfo
  {volume} {91}},\ \bibinfo {pages} {043902} (\bibinfo {year}
  {2003})}\BibitemShut {NoStop}%
\bibitem [{\citenamefont {Aspelmeyer}\ \emph {et~al.}(2014)\citenamefont
  {Aspelmeyer}, \citenamefont {Kippenberg},\ and\ \citenamefont
  {Marquardt}}]{aspelmeyerCavityOptomechanics2014}%
  \BibitemOpen
  \bibfield  {author} {\bibinfo {author} {\bibfnamefont {M.}~\bibnamefont
  {Aspelmeyer}}, \bibinfo {author} {\bibfnamefont {T.~J.}\ \bibnamefont
  {Kippenberg}}, \ and\ \bibinfo {author} {\bibfnamefont {F.}~\bibnamefont
  {Marquardt}},\ }\href {\doibase 10.1103/RevModPhys.86.1391} {\bibfield
  {journal} {\bibinfo  {journal} {Rev. Mod. Phys.}\ }\textbf {\bibinfo {volume}
  {86}},\ \bibinfo {pages} {1391} (\bibinfo {year} {2014})}\BibitemShut
  {NoStop}%
\bibitem [{\citenamefont {Bahl}\ \emph {et~al.}(2011)\citenamefont {Bahl},
  \citenamefont {Zehnpfennig}, \citenamefont {Tomes},\ and\ \citenamefont
  {Carmon}}]{bahlStimulatedOptomechanical2011}%
  \BibitemOpen
  \bibfield  {author} {\bibinfo {author} {\bibfnamefont {G.}~\bibnamefont
  {Bahl}}, \bibinfo {author} {\bibfnamefont {J.}~\bibnamefont {Zehnpfennig}},
  \bibinfo {author} {\bibfnamefont {M.}~\bibnamefont {Tomes}}, \ and\ \bibinfo
  {author} {\bibfnamefont {T.}~\bibnamefont {Carmon}},\ }\href {\doibase
  10.1038/ncomms1412} {\bibfield  {journal} {\bibinfo  {journal} {Nat Commun}\
  }\textbf {\bibinfo {volume} {2}},\ \bibinfo {pages} {403} (\bibinfo {year}
  {2011})}\BibitemShut {NoStop}%
\bibitem [{\citenamefont {Kippenberg}\ \emph {et~al.}(2004)\citenamefont
  {Kippenberg}, \citenamefont {Spillane}, \citenamefont {Armani},\ and\
  \citenamefont {Vahala}}]{kippenbeUltralowthre2004}%
  \BibitemOpen
  \bibfield  {author} {\bibinfo {author} {\bibfnamefont {T.~J.}\ \bibnamefont
  {Kippenberg}}, \bibinfo {author} {\bibfnamefont {S.~M.}\ \bibnamefont
  {Spillane}}, \bibinfo {author} {\bibfnamefont {D.~K.}\ \bibnamefont
  {Armani}}, \ and\ \bibinfo {author} {\bibfnamefont {K.~J.}\ \bibnamefont
  {Vahala}},\ }\href {\doibase 10.1364/OL.29.001224} {\bibfield  {journal}
  {\bibinfo  {journal} {Optics Letters}\ }\textbf {\bibinfo {volume} {29}},\
  \bibinfo {pages} {1224} (\bibinfo {year} {2004})}\BibitemShut {NoStop}%
\bibitem [{\citenamefont {Ilchenko}\ \emph {et~al.}(2004)\citenamefont
  {Ilchenko}, \citenamefont {Savchenkov}, \citenamefont {Matsko},\ and\
  \citenamefont {Maleki}}]{ilchenkoNonlinearOp2004}%
  \BibitemOpen
  \bibfield  {author} {\bibinfo {author} {\bibfnamefont {V.~S.}\ \bibnamefont
  {Ilchenko}}, \bibinfo {author} {\bibfnamefont {A.~A.}\ \bibnamefont
  {Savchenkov}}, \bibinfo {author} {\bibfnamefont {A.~B.}\ \bibnamefont
  {Matsko}}, \ and\ \bibinfo {author} {\bibfnamefont {L.}~\bibnamefont
  {Maleki}},\ }\href {\doibase 10.1103/PhysRevLett.92.043903} {\bibfield
  {journal} {\bibinfo  {journal} {Physical Review Letters}\ }\textbf {\bibinfo
  {volume} {92}},\ \bibinfo {pages} {043903} (\bibinfo {year}
  {2004})}\BibitemShut {NoStop}%
\bibitem [{\citenamefont {Lin}\ \emph {et~al.}(2017)\citenamefont {Lin},
  \citenamefont {Coillet},\ and\ \citenamefont {Chembo}}]{linNonlinearPh2017}%
  \BibitemOpen
  \bibfield  {author} {\bibinfo {author} {\bibfnamefont {G.}~\bibnamefont
  {Lin}}, \bibinfo {author} {\bibfnamefont {A.}~\bibnamefont {Coillet}}, \ and\
  \bibinfo {author} {\bibfnamefont {Y.~K.}\ \bibnamefont {Chembo}},\ }\href
  {\doibase 10.1364/AOP.9.000828} {\bibfield  {journal} {\bibinfo  {journal}
  {Advances in Optics and Photonics}\ }\textbf {\bibinfo {volume} {9}},\
  \bibinfo {pages} {828} (\bibinfo {year} {2017})}\BibitemShut {NoStop}%
\bibitem [{\citenamefont {Chembo}(2016)}]{chemboKerrOptical2016}%
  \BibitemOpen
  \bibfield  {author} {\bibinfo {author} {\bibfnamefont {Y.~K.}\ \bibnamefont
  {Chembo}},\ }\href {\doibase 10.1515/nanoph-2016-0013} {\bibfield  {journal}
  {\bibinfo  {journal} {Nanophotonics}\ }\textbf {\bibinfo {volume} {5}},\
  \bibinfo {pages} {214} (\bibinfo {year} {2016})}\BibitemShut {NoStop}%
\bibitem [{\citenamefont {Del'Haye}\ \emph {et~al.}(2007)\citenamefont
  {Del'Haye}, \citenamefont {Schliesser}, \citenamefont {Arcizet},
  \citenamefont {Wilken}, \citenamefont {Holzwarth},\ and\ \citenamefont
  {Kippenberg}}]{delhayeOpticalFreq2007}%
  \BibitemOpen
  \bibfield  {author} {\bibinfo {author} {\bibfnamefont {P.}~\bibnamefont
  {Del'Haye}}, \bibinfo {author} {\bibfnamefont {A.}~\bibnamefont
  {Schliesser}}, \bibinfo {author} {\bibfnamefont {O.}~\bibnamefont {Arcizet}},
  \bibinfo {author} {\bibfnamefont {T.}~\bibnamefont {Wilken}}, \bibinfo
  {author} {\bibfnamefont {R.}~\bibnamefont {Holzwarth}}, \ and\ \bibinfo
  {author} {\bibfnamefont {T.~J.}\ \bibnamefont {Kippenberg}},\ }\href
  {\doibase 10.1038/nature06401} {\bibfield  {journal} {\bibinfo  {journal}
  {Nature}\ }\textbf {\bibinfo {volume} {450}},\ \bibinfo {pages} {1214}
  (\bibinfo {year} {2007})}\BibitemShut {NoStop}%
\bibitem [{\citenamefont {Kippenberg}\ \emph {et~al.}(2011)\citenamefont
  {Kippenberg}, \citenamefont {Holzwarth},\ and\ \citenamefont
  {Diddams}}]{kippenbeMicroresonat2011}%
  \BibitemOpen
  \bibfield  {author} {\bibinfo {author} {\bibfnamefont {T.~J.}\ \bibnamefont
  {Kippenberg}}, \bibinfo {author} {\bibfnamefont {R.}~\bibnamefont
  {Holzwarth}}, \ and\ \bibinfo {author} {\bibfnamefont {S.~A.}\ \bibnamefont
  {Diddams}},\ }\href {\doibase 10.1126/science.1193968} {\bibfield  {journal}
  {\bibinfo  {journal} {Science}\ }\textbf {\bibinfo {volume} {332}},\ \bibinfo
  {pages} {555} (\bibinfo {year} {2011})}\BibitemShut {NoStop}%
\bibitem [{\citenamefont {Chembo}\ and\ \citenamefont
  {Yu}(2010)}]{chemboModalExpans2010}%
  \BibitemOpen
  \bibfield  {author} {\bibinfo {author} {\bibfnamefont {Y.~K.}\ \bibnamefont
  {Chembo}}\ and\ \bibinfo {author} {\bibfnamefont {N.}~\bibnamefont {Yu}},\
  }\href {\doibase 10.1103/PhysRevA.82.033801} {\bibfield  {journal} {\bibinfo
  {journal} {Physical Review A}\ }\textbf {\bibinfo {volume} {82}},\ \bibinfo
  {pages} {033801} (\bibinfo {year} {2010})}\BibitemShut {NoStop}%
\bibitem [{\citenamefont {Lugiato}\ and\ \citenamefont
  {Lefever}(1987)}]{lugiatoSpatialDissipative1987}%
  \BibitemOpen
  \bibfield  {author} {\bibinfo {author} {\bibfnamefont {L.~A.}\ \bibnamefont
  {Lugiato}}\ and\ \bibinfo {author} {\bibfnamefont {R.}~\bibnamefont
  {Lefever}},\ }\href {\doibase 10.1103/PhysRevLett.58.2209} {\bibfield
  {journal} {\bibinfo  {journal} {Physical Review Letters}\ }\textbf {\bibinfo
  {volume} {58}},\ \bibinfo {pages} {2209} (\bibinfo {year}
  {1987})}\BibitemShut {NoStop}%
\bibitem [{\citenamefont {Chembo}\ and\ \citenamefont
  {Menyuk}(2013)}]{chemboSpatiotemporalLugiatoLefever2013}%
  \BibitemOpen
  \bibfield  {author} {\bibinfo {author} {\bibfnamefont {Y.~K.}\ \bibnamefont
  {Chembo}}\ and\ \bibinfo {author} {\bibfnamefont {C.~R.}\ \bibnamefont
  {Menyuk}},\ }\href {\doibase 10.1103/PhysRevA.87.053852} {\bibfield
  {journal} {\bibinfo  {journal} {Phys. Rev. A}\ }\textbf {\bibinfo {volume}
  {87}},\ \bibinfo {pages} {053852} (\bibinfo {year} {2013})}\BibitemShut
  {NoStop}%
\bibitem [{\citenamefont {Coen}\ \emph {et~al.}(2013)\citenamefont {Coen},
  \citenamefont {Randle}, \citenamefont {Sylvestre},\ and\ \citenamefont
  {Erkintalo}}]{coenModelingOctavespanning2013}%
  \BibitemOpen
  \bibfield  {author} {\bibinfo {author} {\bibfnamefont {S.}~\bibnamefont
  {Coen}}, \bibinfo {author} {\bibfnamefont {H.~G.}\ \bibnamefont {Randle}},
  \bibinfo {author} {\bibfnamefont {T.}~\bibnamefont {Sylvestre}}, \ and\
  \bibinfo {author} {\bibfnamefont {M.}~\bibnamefont {Erkintalo}},\ }\href
  {\doibase 10.1364/OL.38.000037} {\bibfield  {journal} {\bibinfo  {journal}
  {Opt. Lett.}\ }\textbf {\bibinfo {volume} {38}},\ \bibinfo {pages} {37}
  (\bibinfo {year} {2013})}\BibitemShut {NoStop}%
\bibitem [{\citenamefont {Herr}\ \emph {et~al.}(2013)\citenamefont {Herr},
  \citenamefont {Brasch}, \citenamefont {Jost}, \citenamefont {Wang},
  \citenamefont {Kondratiev}, \citenamefont {Gorodetsky},\ and\ \citenamefont
  {Kippenberg}}]{herrTemporalSol2013}%
  \BibitemOpen
  \bibfield  {author} {\bibinfo {author} {\bibfnamefont {T.}~\bibnamefont
  {Herr}}, \bibinfo {author} {\bibfnamefont {V.}~\bibnamefont {Brasch}},
  \bibinfo {author} {\bibfnamefont {J.~D.}\ \bibnamefont {Jost}}, \bibinfo
  {author} {\bibfnamefont {C.~Y.}\ \bibnamefont {Wang}}, \bibinfo {author}
  {\bibfnamefont {N.~M.}\ \bibnamefont {Kondratiev}}, \bibinfo {author}
  {\bibfnamefont {M.~L.}\ \bibnamefont {Gorodetsky}}, \ and\ \bibinfo {author}
  {\bibfnamefont {T.~J.}\ \bibnamefont {Kippenberg}},\ }\href {\doibase
  10.1038/nphoton.2013.343} {\bibfield  {journal} {\bibinfo  {journal} {Nature
  Photonics}\ }\textbf {\bibinfo {volume} {8}},\ \bibinfo {pages} {145}
  (\bibinfo {year} {2013})}\BibitemShut {NoStop}%
\bibitem [{\citenamefont {Kippenberg}\ \emph {et~al.}(2018)\citenamefont
  {Kippenberg}, \citenamefont {Gaeta}, \citenamefont {Lipson},\ and\
  \citenamefont {Gorodetsky}}]{kippenbergDissipativeKerr2018}%
  \BibitemOpen
  \bibfield  {author} {\bibinfo {author} {\bibfnamefont {T.~J.}\ \bibnamefont
  {Kippenberg}}, \bibinfo {author} {\bibfnamefont {A.~L.}\ \bibnamefont
  {Gaeta}}, \bibinfo {author} {\bibfnamefont {M.}~\bibnamefont {Lipson}}, \
  and\ \bibinfo {author} {\bibfnamefont {M.~L.}\ \bibnamefont {Gorodetsky}},\
  }\href {\doibase 10.1126/science.aan8083} {\bibfield  {journal} {\bibinfo
  {journal} {Science}\ }\textbf {\bibinfo {volume} {361}},\ \bibinfo {pages}
  {eaan8083} (\bibinfo {year} {2018})}\BibitemShut {NoStop}%
\bibitem [{\citenamefont {Zhang}\ \emph {et~al.}(2019)\citenamefont {Zhang},
  \citenamefont {Silver}, \citenamefont {Del~Bino}, \citenamefont {Copie},
  \citenamefont {Woodley}, \citenamefont {Ghalanos}, \citenamefont {Svela},
  \citenamefont {Moroney},\ and\ \citenamefont
  {Del'Haye}}]{zhangSubmilliwattlevelMicroresonator2019}%
  \BibitemOpen
  \bibfield  {author} {\bibinfo {author} {\bibfnamefont {S.}~\bibnamefont
  {Zhang}}, \bibinfo {author} {\bibfnamefont {J.~M.}\ \bibnamefont {Silver}},
  \bibinfo {author} {\bibfnamefont {L.}~\bibnamefont {Del~Bino}}, \bibinfo
  {author} {\bibfnamefont {F.}~\bibnamefont {Copie}}, \bibinfo {author}
  {\bibfnamefont {M.~T.~M.}\ \bibnamefont {Woodley}}, \bibinfo {author}
  {\bibfnamefont {G.~N.}\ \bibnamefont {Ghalanos}}, \bibinfo {author}
  {\bibfnamefont {A.~{\O}.}\ \bibnamefont {Svela}}, \bibinfo {author}
  {\bibfnamefont {N.}~\bibnamefont {Moroney}}, \ and\ \bibinfo {author}
  {\bibfnamefont {P.}~\bibnamefont {Del'Haye}},\ }\href {\doibase
  10.1364/OPTICA.6.000206} {\bibfield  {journal} {\bibinfo  {journal} {Optica}\
  }\textbf {\bibinfo {volume} {6}},\ \bibinfo {pages} {206} (\bibinfo {year}
  {2019})}\BibitemShut {NoStop}%
\bibitem [{\citenamefont {Cao}\ \emph {et~al.}(2017)\citenamefont {Cao},
  \citenamefont {Wang}, \citenamefont {Dong}, \citenamefont {Jing},
  \citenamefont {Liu}, \citenamefont {Chen}, \citenamefont {Ge}, \citenamefont
  {Gong},\ and\ \citenamefont {Xiao}}]{caoExperimentalDemonstration2017}%
  \BibitemOpen
  \bibfield  {author} {\bibinfo {author} {\bibfnamefont {Q.~T.}\ \bibnamefont
  {Cao}}, \bibinfo {author} {\bibfnamefont {H.~M.}\ \bibnamefont {Wang}},
  \bibinfo {author} {\bibfnamefont {C.~H.}\ \bibnamefont {Dong}}, \bibinfo
  {author} {\bibfnamefont {H.}~\bibnamefont {Jing}}, \bibinfo {author}
  {\bibfnamefont {R.~S.}\ \bibnamefont {Liu}}, \bibinfo {author} {\bibfnamefont
  {X.}~\bibnamefont {Chen}}, \bibinfo {author} {\bibfnamefont {L.}~\bibnamefont
  {Ge}}, \bibinfo {author} {\bibfnamefont {Q.~H.}\ \bibnamefont {Gong}}, \ and\
  \bibinfo {author} {\bibfnamefont {Y.~F.}\ \bibnamefont {Xiao}},\ }\href
  {\doibase 10.1103/PhysRevLett.118.033901} {\bibfield  {journal} {\bibinfo
  {journal} {Physical Review Letters}\ }\textbf {\bibinfo {volume} {118}},\
  \bibinfo {pages} {033901} (\bibinfo {year} {2017})}\BibitemShut {NoStop}%
\bibitem [{\citenamefont {Del~Bino}\ \emph {et~al.}(2017)\citenamefont
  {Del~Bino}, \citenamefont {Silver}, \citenamefont {Stebbings},\ and\
  \citenamefont {Del'Haye}}]{delbinoSymmetryBreaking2017}%
  \BibitemOpen
  \bibfield  {author} {\bibinfo {author} {\bibfnamefont {L.}~\bibnamefont
  {Del~Bino}}, \bibinfo {author} {\bibfnamefont {J.~M.}\ \bibnamefont
  {Silver}}, \bibinfo {author} {\bibfnamefont {S.~L.}\ \bibnamefont
  {Stebbings}}, \ and\ \bibinfo {author} {\bibfnamefont {P.}~\bibnamefont
  {Del'Haye}},\ }\href {\doibase 10.1038/srep43142} {\bibfield  {journal}
  {\bibinfo  {journal} {Scientific Reports}\ }\textbf {\bibinfo {volume} {7}},\
  \bibinfo {pages} {43142} (\bibinfo {year} {2017})}\BibitemShut {NoStop}%
\bibitem [{\citenamefont {Woodley}\ \emph {et~al.}(2018)\citenamefont
  {Woodley}, \citenamefont {Silver}, \citenamefont {Hill}, \citenamefont
  {Copie}, \citenamefont {Del~Bino}, \citenamefont {Zhang}, \citenamefont
  {Oppo},\ and\ \citenamefont
  {Del'Haye}}]{woodleyUniversalSymmetrybreaking2018}%
  \BibitemOpen
  \bibfield  {author} {\bibinfo {author} {\bibfnamefont {M.~T.~M.}\
  \bibnamefont {Woodley}}, \bibinfo {author} {\bibfnamefont {J.~M.}\
  \bibnamefont {Silver}}, \bibinfo {author} {\bibfnamefont {L.}~\bibnamefont
  {Hill}}, \bibinfo {author} {\bibfnamefont {F.}~\bibnamefont {Copie}},
  \bibinfo {author} {\bibfnamefont {L.}~\bibnamefont {Del~Bino}}, \bibinfo
  {author} {\bibfnamefont {S.~Y.}\ \bibnamefont {Zhang}}, \bibinfo {author}
  {\bibfnamefont {G.~L.}\ \bibnamefont {Oppo}}, \ and\ \bibinfo {author}
  {\bibfnamefont {P.}~\bibnamefont {Del'Haye}},\ }\href {\doibase
  10.1103/PhysRevA.98.053863} {\bibfield  {journal} {\bibinfo  {journal}
  {Physical Review A}\ }\textbf {\bibinfo {volume} {98}},\ \bibinfo {pages}
  {053863} (\bibinfo {year} {2018})}\BibitemShut {NoStop}%
\bibitem [{\citenamefont {Silver}\ \emph {et~al.}(2019)\citenamefont {Silver},
  \citenamefont {Grattan},\ and\ \citenamefont
  {Del'Haye}}]{silverCriticalDyn2019}%
  \BibitemOpen
  \bibfield  {author} {\bibinfo {author} {\bibfnamefont {J.~M.}\ \bibnamefont
  {Silver}}, \bibinfo {author} {\bibfnamefont {K.~T.}\ \bibnamefont {Grattan}},
  \ and\ \bibinfo {author} {\bibfnamefont {P.}~\bibnamefont {Del'Haye}},\
  }\href@noop {} {\bibfield  {journal} {\bibinfo  {journal} {arXiv preprint
  arXiv:1912.08262}\ } (\bibinfo {year} {2019})}\BibitemShut {NoStop}%
\bibitem [{\citenamefont {Chen}\ \emph {et~al.}(2017)\citenamefont {Chen},
  \citenamefont {{\"O}zdemir}, \citenamefont {Zhao}, \citenamefont {Wiersig},\
  and\ \citenamefont {Yang}}]{chenExceptionalPoints2017}%
  \BibitemOpen
  \bibfield  {author} {\bibinfo {author} {\bibfnamefont {W.}~\bibnamefont
  {Chen}}, \bibinfo {author} {\bibfnamefont {{\c S}.~K.}\ \bibnamefont
  {{\"O}zdemir}}, \bibinfo {author} {\bibfnamefont {G.}~\bibnamefont {Zhao}},
  \bibinfo {author} {\bibfnamefont {J.}~\bibnamefont {Wiersig}}, \ and\
  \bibinfo {author} {\bibfnamefont {L.}~\bibnamefont {Yang}},\ }\href {\doibase
  10.1038/nature23281} {\bibfield  {journal} {\bibinfo  {journal} {Nature}\
  }\textbf {\bibinfo {volume} {548}},\ \bibinfo {pages} {192} (\bibinfo {year}
  {2017})}\BibitemShut {NoStop}%
\bibitem [{\citenamefont {Lai}\ \emph {et~al.}(2019)\citenamefont {Lai},
  \citenamefont {Lu}, \citenamefont {Suh}, \citenamefont {Yuan},\ and\
  \citenamefont {Vahala}}]{laiObservation2019}%
  \BibitemOpen
  \bibfield  {author} {\bibinfo {author} {\bibfnamefont {Y.-H.}\ \bibnamefont
  {Lai}}, \bibinfo {author} {\bibfnamefont {Y.-K.}\ \bibnamefont {Lu}},
  \bibinfo {author} {\bibfnamefont {M.-G.}\ \bibnamefont {Suh}}, \bibinfo
  {author} {\bibfnamefont {Z.}~\bibnamefont {Yuan}}, \ and\ \bibinfo {author}
  {\bibfnamefont {K.}~\bibnamefont {Vahala}},\ }\href {\doibase
  10.1038/s41586-019-1777-z} {\bibfield  {journal} {\bibinfo  {journal}
  {Nature}\ }\textbf {\bibinfo {volume} {576}},\ \bibinfo {pages} {65}
  (\bibinfo {year} {2019})}\BibitemShut {NoStop}%
\bibitem [{\citenamefont {Silver}\ \emph {et~al.}(2021)\citenamefont {Silver},
  \citenamefont {Bino}, \citenamefont {Woodley}, \citenamefont {Ghalanos},
  \citenamefont {Ghalanos}, \citenamefont {Svela}, \citenamefont {Moroney},
  \citenamefont {Zhang}, \citenamefont {Grattan}, \citenamefont {Del'Haye},
  \citenamefont {Del'Haye},\ and\ \citenamefont
  {Del'Haye}}]{silverNonlinearEn2021a}%
  \BibitemOpen
  \bibfield  {author} {\bibinfo {author} {\bibfnamefont {J.~M.}\ \bibnamefont
  {Silver}}, \bibinfo {author} {\bibfnamefont {L.~D.}\ \bibnamefont {Bino}},
  \bibinfo {author} {\bibfnamefont {M.~T.~M.}\ \bibnamefont {Woodley}},
  \bibinfo {author} {\bibfnamefont {G.~N.}\ \bibnamefont {Ghalanos}}, \bibinfo
  {author} {\bibfnamefont {G.~N.}\ \bibnamefont {Ghalanos}}, \bibinfo {author}
  {\bibfnamefont {A.~{\O}.}\ \bibnamefont {Svela}}, \bibinfo {author}
  {\bibfnamefont {N.}~\bibnamefont {Moroney}}, \bibinfo {author} {\bibfnamefont
  {S.}~\bibnamefont {Zhang}}, \bibinfo {author} {\bibfnamefont {K.~T.~V.}\
  \bibnamefont {Grattan}}, \bibinfo {author} {\bibfnamefont {P.}~\bibnamefont
  {Del'Haye}}, \bibinfo {author} {\bibfnamefont {P.}~\bibnamefont {Del'Haye}},
  \ and\ \bibinfo {author} {\bibfnamefont {P.}~\bibnamefont {Del'Haye}},\
  }\href {\doibase 10.1364/OPTICA.426018} {\bibfield  {journal} {\bibinfo
  {journal} {Optica}\ }\textbf {\bibinfo {volume} {8}},\ \bibinfo {pages}
  {1219} (\bibinfo {year} {2021})}\BibitemShut {NoStop}%
\bibitem [{\citenamefont {Wang}\ and\ \citenamefont
  {Search}(2015)}]{wangNonlinearMicroresonator2015}%
  \BibitemOpen
  \bibfield  {author} {\bibinfo {author} {\bibfnamefont {C.}~\bibnamefont
  {Wang}}\ and\ \bibinfo {author} {\bibfnamefont {C.~P.}\ \bibnamefont
  {Search}},\ }\href {\doibase 10.1109/JLT.2015.2464105} {\bibfield  {journal}
  {\bibinfo  {journal} {Journal of Lightwave Technology}\ }\textbf {\bibinfo
  {volume} {33}},\ \bibinfo {pages} {4360} (\bibinfo {year}
  {2015})}\BibitemShut {NoStop}%
\bibitem [{\citenamefont {Del~Bino}\ \emph {et~al.}(2018)\citenamefont
  {Del~Bino}, \citenamefont {Silver}, \citenamefont {Woodley}, \citenamefont
  {Stebbings}, \citenamefont {Zhao},\ and\ \citenamefont
  {Del'Haye}}]{delbinoMicroresonat2018}%
  \BibitemOpen
  \bibfield  {author} {\bibinfo {author} {\bibfnamefont {L.}~\bibnamefont
  {Del~Bino}}, \bibinfo {author} {\bibfnamefont {J.~M.}\ \bibnamefont
  {Silver}}, \bibinfo {author} {\bibfnamefont {M.~T.~M.}\ \bibnamefont
  {Woodley}}, \bibinfo {author} {\bibfnamefont {S.~L.}\ \bibnamefont
  {Stebbings}}, \bibinfo {author} {\bibfnamefont {X.}~\bibnamefont {Zhao}}, \
  and\ \bibinfo {author} {\bibfnamefont {P.}~\bibnamefont {Del'Haye}},\ }\href
  {\doibase 10.1364/OPTICA.5.000279} {\bibfield  {journal} {\bibinfo  {journal}
  {Optica}\ }\textbf {\bibinfo {volume} {5}},\ \bibinfo {pages} {279} (\bibinfo
  {year} {2018})}\BibitemShut {NoStop}%
\bibitem [{\citenamefont {Del~Bino}\ \emph {et~al.}(2021)\citenamefont
  {Del~Bino}, \citenamefont {Moroney},\ and\ \citenamefont
  {Del'Haye}}]{delbinoOpticalMemo2021}%
  \BibitemOpen
  \bibfield  {author} {\bibinfo {author} {\bibfnamefont {L.}~\bibnamefont
  {Del~Bino}}, \bibinfo {author} {\bibfnamefont {N.}~\bibnamefont {Moroney}}, \
  and\ \bibinfo {author} {\bibfnamefont {P.}~\bibnamefont {Del'Haye}},\ }\href
  {\doibase 10.1364/OE.417951} {\bibfield  {journal} {\bibinfo  {journal} {Opt.
  Express}\ }\textbf {\bibinfo {volume} {29}},\ \bibinfo {pages} {2193}
  (\bibinfo {year} {2021})}\BibitemShut {NoStop}%
\bibitem [{\citenamefont {Moroney}\ \emph {et~al.}(2020)\citenamefont
  {Moroney}, \citenamefont {Del~Bino}, \citenamefont {Woodley}, \citenamefont
  {Ghalanos}, \citenamefont {Silver}, \citenamefont {Svela}, \citenamefont
  {Zhang},\ and\ \citenamefont {Del'Haye}}]{moroneyLogicGates2020}%
  \BibitemOpen
  \bibfield  {author} {\bibinfo {author} {\bibfnamefont {N.}~\bibnamefont
  {Moroney}}, \bibinfo {author} {\bibfnamefont {L.}~\bibnamefont {Del~Bino}},
  \bibinfo {author} {\bibfnamefont {M.~T.~M.}\ \bibnamefont {Woodley}},
  \bibinfo {author} {\bibfnamefont {G.~N.}\ \bibnamefont {Ghalanos}}, \bibinfo
  {author} {\bibfnamefont {J.~M.}\ \bibnamefont {Silver}}, \bibinfo {author}
  {\bibfnamefont {A.~O.}\ \bibnamefont {Svela}}, \bibinfo {author}
  {\bibfnamefont {S.}~\bibnamefont {Zhang}}, \ and\ \bibinfo {author}
  {\bibfnamefont {P.}~\bibnamefont {Del'Haye}},\ }\href {\doibase
  10.1109/JLT.2020.2975119} {\bibfield  {journal} {\bibinfo  {journal} {J.
  Lightwave Technol.}\ }\textbf {\bibinfo {volume} {38}},\ \bibinfo {pages}
  {1414} (\bibinfo {year} {2020})}\BibitemShut {NoStop}%
\bibitem [{\citenamefont {Kaplan}\ and\ \citenamefont
  {Meystre}(1981)}]{kaplanEnhancementSagnac1981}%
  \BibitemOpen
  \bibfield  {author} {\bibinfo {author} {\bibfnamefont {A.~E.}\ \bibnamefont
  {Kaplan}}\ and\ \bibinfo {author} {\bibfnamefont {P.}~\bibnamefont
  {Meystre}},\ }\href {\doibase 10.1364/OL.6.000590} {\bibfield  {journal}
  {\bibinfo  {journal} {Optics Letters}\ }\textbf {\bibinfo {volume} {6}},\
  \bibinfo {pages} {590} (\bibinfo {year} {1981})}\BibitemShut {NoStop}%
\bibitem [{\citenamefont {Ghalanos}\ \emph {et~al.}(2020)\citenamefont
  {Ghalanos}, \citenamefont {Silver}, \citenamefont {Del~Bino}, \citenamefont
  {Moroney}, \citenamefont {Zhang}, \citenamefont {Woodley}, \citenamefont
  {Svela},\ and\ \citenamefont
  {Del'Haye}}]{ghalanosKerrNonlinearityInducedModeSplitting2020}%
  \BibitemOpen
  \bibfield  {author} {\bibinfo {author} {\bibfnamefont {G.~N.}\ \bibnamefont
  {Ghalanos}}, \bibinfo {author} {\bibfnamefont {J.~M.}\ \bibnamefont
  {Silver}}, \bibinfo {author} {\bibfnamefont {L.}~\bibnamefont {Del~Bino}},
  \bibinfo {author} {\bibfnamefont {N.}~\bibnamefont {Moroney}}, \bibinfo
  {author} {\bibfnamefont {S.}~\bibnamefont {Zhang}}, \bibinfo {author}
  {\bibfnamefont {M.~T.~M.}\ \bibnamefont {Woodley}}, \bibinfo {author}
  {\bibfnamefont {A.~{\O}.}\ \bibnamefont {Svela}}, \ and\ \bibinfo {author}
  {\bibfnamefont {P.}~\bibnamefont {Del'Haye}},\ }\href {\doibase
  10.1103/PhysRevLett.124.223901} {\bibfield  {journal} {\bibinfo  {journal}
  {Phys. Rev. Lett.}\ }\textbf {\bibinfo {volume} {124}},\ \bibinfo {pages}
  {223901} (\bibinfo {year} {2020})}\BibitemShut {NoStop}%
\bibitem [{\citenamefont {Kaplan}(1981)}]{kaplanOpticalBistability1981}%
  \BibitemOpen
  \bibfield  {author} {\bibinfo {author} {\bibfnamefont {A.~E.}\ \bibnamefont
  {Kaplan}},\ }\href {\doibase 10.1364/OL.6.000360} {\bibfield  {journal}
  {\bibinfo  {journal} {Optics Letters}\ }\textbf {\bibinfo {volume} {6}},\
  \bibinfo {pages} {360} (\bibinfo {year} {1981})}\BibitemShut {NoStop}%
\bibitem [{\citenamefont {Kivshar}\ and\ \citenamefont
  {Pelinovsky}(2000)}]{kivsharSelffocusingTransverse2000}%
  \BibitemOpen
  \bibfield  {author} {\bibinfo {author} {\bibfnamefont {Y.~S.}\ \bibnamefont
  {Kivshar}}\ and\ \bibinfo {author} {\bibfnamefont {D.~E.}\ \bibnamefont
  {Pelinovsky}},\ }\href {\doibase 10.1016/S0370-1573(99)00106-4} {\bibfield
  {journal} {\bibinfo  {journal} {Physics Reports}\ }\textbf {\bibinfo {volume}
  {331}},\ \bibinfo {pages} {117} (\bibinfo {year} {2000})}\BibitemShut
  {NoStop}%
\bibitem [{\citenamefont {Yang}\ \emph {et~al.}(2017)\citenamefont {Yang},
  \citenamefont {Yi}, \citenamefont {Yang},\ and\ \citenamefont
  {Vahala}}]{yangCounterpropagatingSolitons2017}%
  \BibitemOpen
  \bibfield  {author} {\bibinfo {author} {\bibfnamefont {Q.-F.}\ \bibnamefont
  {Yang}}, \bibinfo {author} {\bibfnamefont {X.}~\bibnamefont {Yi}}, \bibinfo
  {author} {\bibfnamefont {K.~Y.}\ \bibnamefont {Yang}}, \ and\ \bibinfo
  {author} {\bibfnamefont {K.}~\bibnamefont {Vahala}},\ }\href {\doibase
  10.1038/nphoton.2017.117} {\bibfield  {journal} {\bibinfo  {journal} {Nature
  Photonics}\ }\textbf {\bibinfo {volume} {11}},\ \bibinfo {pages} {560}
  (\bibinfo {year} {2017})}\BibitemShut {NoStop}%
\bibitem [{\citenamefont {Joshi}\ \emph {et~al.}(2018)\citenamefont {Joshi},
  \citenamefont {Klenner}, \citenamefont {Okawachi}, \citenamefont {Yu},
  \citenamefont {Luke}, \citenamefont {Ji}, \citenamefont {Lipson},\ and\
  \citenamefont {Gaeta}}]{joshiCounterrotat2018}%
  \BibitemOpen
  \bibfield  {author} {\bibinfo {author} {\bibfnamefont {C.}~\bibnamefont
  {Joshi}}, \bibinfo {author} {\bibfnamefont {A.}~\bibnamefont {Klenner}},
  \bibinfo {author} {\bibfnamefont {Y.}~\bibnamefont {Okawachi}}, \bibinfo
  {author} {\bibfnamefont {M.}~\bibnamefont {Yu}}, \bibinfo {author}
  {\bibfnamefont {K.}~\bibnamefont {Luke}}, \bibinfo {author} {\bibfnamefont
  {X.}~\bibnamefont {Ji}}, \bibinfo {author} {\bibfnamefont {M.}~\bibnamefont
  {Lipson}}, \ and\ \bibinfo {author} {\bibfnamefont {A.~L.}\ \bibnamefont
  {Gaeta}},\ }\href {\doibase 10.1364/OL.43.000547} {\bibfield  {journal}
  {\bibinfo  {journal} {Opt. Lett.}\ }\textbf {\bibinfo {volume} {43}},\
  \bibinfo {pages} {547} (\bibinfo {year} {2018})}\BibitemShut {NoStop}%
\bibitem [{\citenamefont {Bao}\ \emph {et~al.}(2019)\citenamefont {Bao},
  \citenamefont {Liao}, \citenamefont {Kordts}, \citenamefont {Zhang},
  \citenamefont {Matsko}, \citenamefont {Karpov}, \citenamefont {Pfeiffer},
  \citenamefont {Xie}, \citenamefont {Cao}, \citenamefont {Almaiman},
  \citenamefont {Tur}, \citenamefont {Kippenberg},\ and\ \citenamefont
  {Willner}}]{baoOrthogonally2019}%
  \BibitemOpen
  \bibfield  {author} {\bibinfo {author} {\bibfnamefont {C.}~\bibnamefont
  {Bao}}, \bibinfo {author} {\bibfnamefont {P.}~\bibnamefont {Liao}}, \bibinfo
  {author} {\bibfnamefont {A.}~\bibnamefont {Kordts}}, \bibinfo {author}
  {\bibfnamefont {L.}~\bibnamefont {Zhang}}, \bibinfo {author} {\bibfnamefont
  {A.}~\bibnamefont {Matsko}}, \bibinfo {author} {\bibfnamefont
  {M.}~\bibnamefont {Karpov}}, \bibinfo {author} {\bibfnamefont {M.~H.~P.}\
  \bibnamefont {Pfeiffer}}, \bibinfo {author} {\bibfnamefont {G.}~\bibnamefont
  {Xie}}, \bibinfo {author} {\bibfnamefont {Y.}~\bibnamefont {Cao}}, \bibinfo
  {author} {\bibfnamefont {A.}~\bibnamefont {Almaiman}}, \bibinfo {author}
  {\bibfnamefont {M.}~\bibnamefont {Tur}}, \bibinfo {author} {\bibfnamefont
  {T.~J.}\ \bibnamefont {Kippenberg}}, \ and\ \bibinfo {author} {\bibfnamefont
  {A.~E.}\ \bibnamefont {Willner}},\ }\href {\doibase 10.1364/OL.44.001472}
  {\bibfield  {journal} {\bibinfo  {journal} {Opt. Lett.}\ }\textbf {\bibinfo
  {volume} {44}},\ \bibinfo {pages} {1472} (\bibinfo {year}
  {2019})}\BibitemShut {NoStop}%
\bibitem [{\citenamefont {Fatome}\ \emph {et~al.}(2020)\citenamefont {Fatome},
  \citenamefont {Fatome}, \citenamefont {Fatome}, \citenamefont {Kibler},
  \citenamefont {Leo}, \citenamefont {Bendahmane}, \citenamefont {Oppo},
  \citenamefont {Garbin}, \citenamefont {Garbin}, \citenamefont {Garbin},
  \citenamefont {Murdoch}, \citenamefont {Murdoch}, \citenamefont {Erkintalo},
  \citenamefont {Erkintalo}, \citenamefont {Coen},\ and\ \citenamefont
  {Coen}}]{fatomePolarization2020}%
  \BibitemOpen
  \bibfield  {author} {\bibinfo {author} {\bibfnamefont {J.}~\bibnamefont
  {Fatome}}, \bibinfo {author} {\bibfnamefont {J.}~\bibnamefont {Fatome}},
  \bibinfo {author} {\bibfnamefont {J.}~\bibnamefont {Fatome}}, \bibinfo
  {author} {\bibfnamefont {B.}~\bibnamefont {Kibler}}, \bibinfo {author}
  {\bibfnamefont {F.}~\bibnamefont {Leo}}, \bibinfo {author} {\bibfnamefont
  {A.}~\bibnamefont {Bendahmane}}, \bibinfo {author} {\bibfnamefont {G.-L.}\
  \bibnamefont {Oppo}}, \bibinfo {author} {\bibfnamefont {B.}~\bibnamefont
  {Garbin}}, \bibinfo {author} {\bibfnamefont {B.}~\bibnamefont {Garbin}},
  \bibinfo {author} {\bibfnamefont {B.}~\bibnamefont {Garbin}}, \bibinfo
  {author} {\bibfnamefont {S.~G.}\ \bibnamefont {Murdoch}}, \bibinfo {author}
  {\bibfnamefont {S.~G.}\ \bibnamefont {Murdoch}}, \bibinfo {author}
  {\bibfnamefont {M.}~\bibnamefont {Erkintalo}}, \bibinfo {author}
  {\bibfnamefont {M.}~\bibnamefont {Erkintalo}}, \bibinfo {author}
  {\bibfnamefont {S.}~\bibnamefont {Coen}}, \ and\ \bibinfo {author}
  {\bibfnamefont {S.}~\bibnamefont {Coen}},\ }\href {\doibase
  10.1364/OL.400474} {\bibfield  {journal} {\bibinfo  {journal} {Opt. Lett.}\
  }\textbf {\bibinfo {volume} {45}},\ \bibinfo {pages} {5069} (\bibinfo {year}
  {2020})}\BibitemShut {NoStop}%
\bibitem [{\citenamefont {Garbin}\ \emph {et~al.}(2020)\citenamefont {Garbin},
  \citenamefont {Fatome}, \citenamefont {Oppo}, \citenamefont {Erkintalo},
  \citenamefont {Murdoch},\ and\ \citenamefont
  {Coen}}]{garbinAsymmetricBalance2020}%
  \BibitemOpen
  \bibfield  {author} {\bibinfo {author} {\bibfnamefont {B.}~\bibnamefont
  {Garbin}}, \bibinfo {author} {\bibfnamefont {J.}~\bibnamefont {Fatome}},
  \bibinfo {author} {\bibfnamefont {G.-L.}\ \bibnamefont {Oppo}}, \bibinfo
  {author} {\bibfnamefont {M.}~\bibnamefont {Erkintalo}}, \bibinfo {author}
  {\bibfnamefont {S.~G.}\ \bibnamefont {Murdoch}}, \ and\ \bibinfo {author}
  {\bibfnamefont {S.}~\bibnamefont {Coen}},\ }\href {\doibase
  10.1103/PhysRevResearch.2.023244} {\bibfield  {journal} {\bibinfo  {journal}
  {Phys. Rev. Research}\ }\textbf {\bibinfo {volume} {2}},\ \bibinfo {pages}
  {023244} (\bibinfo {year} {2020})}\BibitemShut {NoStop}%
\bibitem [{\citenamefont {Garbin}\ \emph {et~al.}(2021)\citenamefont {Garbin},
  \citenamefont {Fatome}, \citenamefont {Oppo}, \citenamefont {Erkintalo},
  \citenamefont {Murdoch},\ and\ \citenamefont
  {Coen}}]{garbinDissipativePolarization2021}%
  \BibitemOpen
  \bibfield  {author} {\bibinfo {author} {\bibfnamefont {B.}~\bibnamefont
  {Garbin}}, \bibinfo {author} {\bibfnamefont {J.}~\bibnamefont {Fatome}},
  \bibinfo {author} {\bibfnamefont {G.-L.}\ \bibnamefont {Oppo}}, \bibinfo
  {author} {\bibfnamefont {M.}~\bibnamefont {Erkintalo}}, \bibinfo {author}
  {\bibfnamefont {S.~G.}\ \bibnamefont {Murdoch}}, \ and\ \bibinfo {author}
  {\bibfnamefont {S.}~\bibnamefont {Coen}},\ }\href {\doibase
  10.1103/PhysRevLett.126.023904} {\bibfield  {journal} {\bibinfo  {journal}
  {Phys. Rev. Lett.}\ }\textbf {\bibinfo {volume} {126}},\ \bibinfo {pages}
  {023904} (\bibinfo {year} {2021})}\BibitemShut {NoStop}%
\bibitem [{\citenamefont {Hill}\ \emph
  {et~al.}(2020{\natexlab{a}})\citenamefont {Hill}, \citenamefont {Oppo},
  \citenamefont {Woodley},\ and\ \citenamefont
  {Del'Haye}}]{hillEffectsSelf2020}%
  \BibitemOpen
  \bibfield  {author} {\bibinfo {author} {\bibfnamefont {L.}~\bibnamefont
  {Hill}}, \bibinfo {author} {\bibfnamefont {G.-L.}\ \bibnamefont {Oppo}},
  \bibinfo {author} {\bibfnamefont {M.~T.~M.}\ \bibnamefont {Woodley}}, \ and\
  \bibinfo {author} {\bibfnamefont {P.}~\bibnamefont {Del'Haye}},\ }\href
  {\doibase 10.1103/PhysRevA.101.013823} {\bibfield  {journal} {\bibinfo
  {journal} {Phys. Rev. A}\ }\textbf {\bibinfo {volume} {101}},\ \bibinfo
  {pages} {013823} (\bibinfo {year} {2020}{\natexlab{a}})}\BibitemShut
  {NoStop}%
\bibitem [{\citenamefont {Woodley}\ \emph {et~al.}(2021)\citenamefont
  {Woodley}, \citenamefont {Hill}, \citenamefont {Del~Bino}, \citenamefont
  {Oppo},\ and\ \citenamefont {Del'Haye}}]{woodleySelfSwitchingKerr2021}%
  \BibitemOpen
  \bibfield  {author} {\bibinfo {author} {\bibfnamefont {M.~T.~M.}\
  \bibnamefont {Woodley}}, \bibinfo {author} {\bibfnamefont {L.}~\bibnamefont
  {Hill}}, \bibinfo {author} {\bibfnamefont {L.}~\bibnamefont {Del~Bino}},
  \bibinfo {author} {\bibfnamefont {G.-L.}\ \bibnamefont {Oppo}}, \ and\
  \bibinfo {author} {\bibfnamefont {P.}~\bibnamefont {Del'Haye}},\ }\href
  {\doibase 10.1103/PhysRevLett.126.043901} {\bibfield  {journal} {\bibinfo
  {journal} {Phys. Rev. Lett.}\ }\textbf {\bibinfo {volume} {126}},\ \bibinfo
  {pages} {043901} (\bibinfo {year} {2021})}\BibitemShut {NoStop}%
\bibitem [{\citenamefont {Hill}\ \emph
  {et~al.}(2020{\natexlab{b}})\citenamefont {Hill}, \citenamefont {Xu},
  \citenamefont {Fatome}, \citenamefont {Oppo}, \citenamefont {Murdoch},
  \citenamefont {Erkintalo},\ and\ \citenamefont {Coen}}]{hillBreathingDy2020}%
  \BibitemOpen
  \bibfield  {author} {\bibinfo {author} {\bibfnamefont {L.}~\bibnamefont
  {Hill}}, \bibinfo {author} {\bibfnamefont {G.}~\bibnamefont {Xu}}, \bibinfo
  {author} {\bibfnamefont {J.}~\bibnamefont {Fatome}}, \bibinfo {author}
  {\bibfnamefont {G.-L.}\ \bibnamefont {Oppo}}, \bibinfo {author}
  {\bibfnamefont {S.~G.}\ \bibnamefont {Murdoch}}, \bibinfo {author}
  {\bibfnamefont {M.}~\bibnamefont {Erkintalo}}, \ and\ \bibinfo {author}
  {\bibfnamefont {S.}~\bibnamefont {Coen}},\ }in\ \href {\doibase
  10.1364/NP.2020.NpTu1D.3} {\emph {\bibinfo {booktitle} {{{OSA Advanced
  Photonics Congress}} ({{AP}}) 2020 ({{IPR}}, {{NP}}, {{NOMA}}, {{Networks}},
  {{PVLED}}, {{PSC}}, {{SPPCom}}, {{SOF}}) (2020), Paper {{NpTu1D}}.3}}}\
  (\bibinfo  {publisher} {{Optical Society of America}},\ \bibinfo {year}
  {2020})\ p.\ \bibinfo {pages} {NpTu1D.3}\BibitemShut {NoStop}%
\bibitem [{\citenamefont {Xu}\ \emph {et~al.}(2021)\citenamefont {Xu},
  \citenamefont {Nielsen}, \citenamefont {Garbin}, \citenamefont {Hill},
  \citenamefont {Oppo}, \citenamefont {Fatome}, \citenamefont {Murdoch},
  \citenamefont {Coen},\ and\ \citenamefont {Erkintalo}}]{xuSpontaneous2021}%
  \BibitemOpen
  \bibfield  {author} {\bibinfo {author} {\bibfnamefont {G.}~\bibnamefont
  {Xu}}, \bibinfo {author} {\bibfnamefont {A.~U.}\ \bibnamefont {Nielsen}},
  \bibinfo {author} {\bibfnamefont {B.}~\bibnamefont {Garbin}}, \bibinfo
  {author} {\bibfnamefont {L.}~\bibnamefont {Hill}}, \bibinfo {author}
  {\bibfnamefont {G.-L.}\ \bibnamefont {Oppo}}, \bibinfo {author}
  {\bibfnamefont {J.}~\bibnamefont {Fatome}}, \bibinfo {author} {\bibfnamefont
  {S.~G.}\ \bibnamefont {Murdoch}}, \bibinfo {author} {\bibfnamefont
  {S.}~\bibnamefont {Coen}}, \ and\ \bibinfo {author} {\bibfnamefont
  {M.}~\bibnamefont {Erkintalo}},\ }\href {\doibase 10.1038/s41467-021-24251-0}
  {\bibfield  {journal} {\bibinfo  {journal} {Nat Commun}\ }\textbf {\bibinfo
  {volume} {12}},\ \bibinfo {pages} {4023} (\bibinfo {year}
  {2021})}\BibitemShut {NoStop}%
\bibitem [{\citenamefont {Yariv}(2000)}]{yarivUniversalRe2000}%
  \BibitemOpen
  \bibfield  {author} {\bibinfo {author} {\bibfnamefont {A.}~\bibnamefont
  {Yariv}},\ }\href {\doibase 10.1049/el:20000340} {\bibfield  {journal}
  {\bibinfo  {journal} {Electronics Letters}\ }\textbf {\bibinfo {volume}
  {36}},\ \bibinfo {pages} {321} (\bibinfo {year} {2000})}\BibitemShut
  {NoStop}%
\bibitem [{\citenamefont {Gorodetsky}\ \emph {et~al.}(2000)\citenamefont
  {Gorodetsky}, \citenamefont {Pryamikov},\ and\ \citenamefont
  {Ilchenko}}]{gorodetsRayleighSca2000}%
  \BibitemOpen
  \bibfield  {author} {\bibinfo {author} {\bibfnamefont {M.~L.}\ \bibnamefont
  {Gorodetsky}}, \bibinfo {author} {\bibfnamefont {A.~D.}\ \bibnamefont
  {Pryamikov}}, \ and\ \bibinfo {author} {\bibfnamefont {V.~S.}\ \bibnamefont
  {Ilchenko}},\ }\href {\doibase 10.1364/JOSAB.17.001051} {\bibfield  {journal}
  {\bibinfo  {journal} {J. Opt. Soc. Am. B}\ }\textbf {\bibinfo {volume}
  {17}},\ \bibinfo {pages} {1051} (\bibinfo {year} {2000})}\BibitemShut
  {NoStop}%
\bibitem [{\citenamefont {Gorodetsky}\ and\ \citenamefont
  {Fomin}(2007)}]{gorodetsEigenfrequen2007}%
  \BibitemOpen
  \bibfield  {author} {\bibinfo {author} {\bibfnamefont {M.~L.}\ \bibnamefont
  {Gorodetsky}}\ and\ \bibinfo {author} {\bibfnamefont {A.~E.}\ \bibnamefont
  {Fomin}},\ }\href {\doibase 10.1070/QE2007v037n02ABEH013306} {\bibfield
  {journal} {\bibinfo  {journal} {Quantum Electronics}\ }\textbf {\bibinfo
  {volume} {37}},\ \bibinfo {pages} {167} (\bibinfo {year} {2007})}\BibitemShut
  {NoStop}%
\bibitem [{\citenamefont {Lacey}\ and\ \citenamefont
  {Payne}(1990)}]{laceyRadiationLo1990}%
  \BibitemOpen
  \bibfield  {author} {\bibinfo {author} {\bibfnamefont {J.~P.~R.}\
  \bibnamefont {Lacey}}\ and\ \bibinfo {author} {\bibfnamefont {F.~P.}\
  \bibnamefont {Payne}},\ }\href {\doibase 10.1049/ip-j.1990.0047} {\bibfield
  {journal} {\bibinfo  {journal} {IEE Proceedings J (Optoelectronics)}\
  }\textbf {\bibinfo {volume} {137}},\ \bibinfo {pages} {282} (\bibinfo {year}
  {1990})}\BibitemShut {NoStop}%
\bibitem [{\citenamefont {Little}\ and\ \citenamefont
  {Chu}(1996)}]{littleEstimatingS1996}%
  \BibitemOpen
  \bibfield  {author} {\bibinfo {author} {\bibfnamefont {B.~E.}\ \bibnamefont
  {Little}}\ and\ \bibinfo {author} {\bibfnamefont {S.~T.}\ \bibnamefont
  {Chu}},\ }\href {\doibase 10.1364/OL.21.001390} {\bibfield  {journal}
  {\bibinfo  {journal} {Opt. Lett.}\ }\textbf {\bibinfo {volume} {21}},\
  \bibinfo {pages} {1390} (\bibinfo {year} {1996})}\BibitemShut {NoStop}%
\bibitem [{\citenamefont {{\v C}tyrok{\'y}}\ \emph {et~al.}(2006)\citenamefont
  {{\v C}tyrok{\'y}}, \citenamefont {Richter},\ and\ \citenamefont {{\v S}i{\v
  n}or}}]{ctyrokyDualResonan2006}%
  \BibitemOpen
  \bibfield  {author} {\bibinfo {author} {\bibfnamefont {J.}~\bibnamefont {{\v
  C}tyrok{\'y}}}, \bibinfo {author} {\bibfnamefont {I.}~\bibnamefont
  {Richter}}, \ and\ \bibinfo {author} {\bibfnamefont {M.}~\bibnamefont {{\v
  S}i{\v n}or}},\ }\href {\doibase 10.1007/s11082-006-9037-5} {\bibfield
  {journal} {\bibinfo  {journal} {Opt Quant Electron}\ }\textbf {\bibinfo
  {volume} {38}},\ \bibinfo {pages} {781} (\bibinfo {year} {2006})}\BibitemShut
  {NoStop}%
\bibitem [{\citenamefont {Barber}\ and\ \citenamefont
  {Yeh}(1975)}]{barberScatteringE1975}%
  \BibitemOpen
  \bibfield  {author} {\bibinfo {author} {\bibfnamefont {P.}~\bibnamefont
  {Barber}}\ and\ \bibinfo {author} {\bibfnamefont {C.}~\bibnamefont {Yeh}},\
  }\href {\doibase 10.1364/AO.14.002864} {\bibfield  {journal} {\bibinfo
  {journal} {Appl. Opt.}\ }\textbf {\bibinfo {volume} {14}},\ \bibinfo {pages}
  {2864} (\bibinfo {year} {1975})}\BibitemShut {NoStop}%
\bibitem [{\citenamefont {Drummond}\ and\ \citenamefont
  {Hillery}(2014)}]{drummondQuantumTheo2014}%
  \BibitemOpen
  \bibfield  {author} {\bibinfo {author} {\bibfnamefont {P.~D.}\ \bibnamefont
  {Drummond}}\ and\ \bibinfo {author} {\bibfnamefont {M.}~\bibnamefont
  {Hillery}},\ }\href@noop {} {\emph {\bibinfo {title} {The {{Quantum Theory}}
  of {{Nonlinear Optics}}}}}\ (\bibinfo  {publisher} {{Cambridge University
  Press}},\ \bibinfo {year} {2014})\BibitemShut {NoStop}%
\bibitem [{\citenamefont {New}(2011)}]{newIntroduction2011}%
  \BibitemOpen
  \bibfield  {author} {\bibinfo {author} {\bibfnamefont {G.}~\bibnamefont
  {New}},\ }\href@noop {} {\emph {\bibinfo {title} {Introduction to {{Nonlinear
  Optics}}}}}\ (\bibinfo  {publisher} {{Cambridge University Press}},\ \bibinfo
  {year} {2011})\BibitemShut {NoStop}%
\bibitem [{\citenamefont {Bures}(2009)}]{buresGuidedOptic2009}%
  \BibitemOpen
  \bibfield  {author} {\bibinfo {author} {\bibfnamefont {J.}~\bibnamefont
  {Bures}},\ }\href@noop {} {\emph {\bibinfo {title} {Guided {{Optics}}:
  Optical {{Fibers}} and {{All}}-Fiber {{Components}}}}}\ (\bibinfo
  {publisher} {{John Wiley \& Sons}},\ \bibinfo {year} {2009})\BibitemShut
  {NoStop}%
\end{thebibliography}
\end{document}